\documentclass[
bibnotes,
amsmath,amssymb,
aps,
pra,
]{revtex4-2}

\usepackage{graphicx}
\usepackage{dcolumn}
\usepackage{bm}
\usepackage{ulem}[normalem]
\usepackage{xcolor}
\usepackage[english]{babel}
\babeladjust{autoload.bcp47 = on}

\begin{document}



\title{The evolution of a supermassive binary black hole in an non-spherical nuclear star cluster}
\author{Pavel B. Ivanov}
\email{pavel000astrophysics@gmail.com}
\affiliation{%
P.N. Lebedev Physical Institute, 53 Leninsky Prospect, Moscow, 119991, Russia
}%
\author{Alexander G. Polnarev}
\email{a.g.polnarev@qmul.ac.uk}
\affiliation{Queen Mary University of London, Mile End Road, London, E1 4NS,  
UK}%

\begin{abstract}
{ In this Paper we consider a secular orbital evolution of a supermassive binary black hole (SBBH) with unequal masses  $M_p$ and $M_s < M_p$ immersed in a central part of a non-spherical nuclear star cluster (NSC). When  the                                       mass of NSC enclosed inside the orbit 
becomes smaller than $M_{s}$ 
 dynamical friction becomes inefficient. 
The subsequent orbital evolution of SBBH
is largely governed by  perturbing tidal potential of NSC arising from 
its non-sphericity. When the perturbing potential is mainly determined by quadrupole harmonics
with azimuthal number $|m|=2$
the corresponding secular dynamics of the SBBH does not conserve any 
components of the angular momentum  and can lead, for this reason, to the formation of highly eccentric orbits. Then such orbits can experience an efficient circularization  due to emission of gravitational waves (GW). In this Paper we consider this situation in some detail. 

We study analytically the secular evolution 
taking into account the important effect of Einstein apsidal precession 
and estimate the largest possible value of $e$, which can be obtained. 
We confirm and extend our analytic results
by solving numerically the secular equations. 

These results are used to estimate a possibility of fast orbital circularization through emission of
gravitation waves on a highly eccentric orbit,  with circularization timescale of the
order of the orbital period. We find that our mechanism could result in such events on a time scale of the order of or smaller than a few Gyr.
It is stressed that for particular values of the model parameters such events may sometimes  be distinguished from the ones expected in more
standard scenarios, since in our case the eccentricity may remain substantial all the way down to the
final merger. 

It is also noted that our results can be applied to other astrophysical settings, e.g. to study
the orbital evolution of a binary star or a proto planetary system inside a massive deformed gas cloud.} 

\end{abstract}

\maketitle


\onecolumngrid
\section{Introduction}

The possibility that supermassive binary black holes (SBBHs) may harbor in centers of many galaxies is a natural consequence of the process of galaxy mergers and the subsequent evolution of two gravitationally unbound black holes toward the center of mass of the newly formed galaxy due to dynamical friction. The idea of the existence of such objects was put forward in several papers; see, for example, \cite{Komb} and \cite{BBR} and was explored later by numerous researchers; see
e.g. \cite{Merrev} for a review of the appropriate theoretical aspects and \cite{Kom2} for a review of potential candidates found in observations (see also \cite{Gurvits}). 
In particular, SBBHs may manifest themselves as sources of various nonstandard activity in galactic centers, with perhaps the most well
known example being the explanation of the quasi-periodic outbursts observed in the quasar OJ 287. 
\cite{Valt}\footnote{Note, however, that the SBBH model of OJ 287
with the standard parameters of the system was recently criticized both from observational and theoretical points of view, see  \cite{Kom1} and
\cite{IZ}, respectively}. Even more importantly, SBBHs may provide the most powerful sources for future space-based gravitational wave antennas,
see, e.g. \cite{Grish} and \cite{Amaro}, \cite{Auc} for a more recent review.

It was pointed out in \cite{BBR} that when a SBBH becomes gravitationally bound its orbital evolution may be so slow that
the formation of the sources of gravitational waves could be impossible for cosmological, i.e. Hubble,  time. This problem was nicknamed as 'the final parsec problem'. 
Over the years several possible solutions to the final parsec problem were proposed, such as e.g. the interaction of SBBH with gaseous environment
(e.g. in the form of an accretion disk, see \cite{IPP}, and also \cite{Zr} for a recent discussion
and references), or triple systems of supermassive black holes, see e.g. \cite{blaes} and \cite{Koz1}, or the possibility of
the central stellar population to have a non-spherical distribution. We
consider below the latter possibility in some detail.

A non-spherical shape of the central regions of a galaxy may alleviate the central parsec problem in two ways. At first, it leads to a larger bulk of 
stars able to gravitationally interact with SBBH due to non-conservation of the absolute values of their angular momenta, thus facilitating the 
process of hardening of a gravitationally bound supermassive binary, see \cite{Mer1}. Secondly, since the angular momentum of the binary is also 
not conserved its orbital evolution could lead, in principle, to the formation of eccentric orbits in a way analogous to the well known 
von Zeipel-Lidov-
Kozai effect \cite{Zeipel}, \cite{Lid} and \cite{Koz}, hereafter ZLK, see \cite{Ito} for a detailed  historical review of these studies. If periastron of such
eccentric orbits is small enough the emission of gravitational waves can lead to an efficient loss of orbital energy with the subsequent merger of 
the black holes in a time, smaller than the Hubble time.

So far, the problem of SBBHs evolution in the central parts of a non-spherical galaxy has been mainly treated by numerical calculations. 
However,  since the numerical approach confronts significant difficulties, some additional simplification have been used in these numerical studies.
In particular, in certain papers the authors assumed that SBBH's orbit is circular, see e.g. \cite{Chen}, or the stellar distribution is either
spherically symmetric, see e.g. \cite{Ses} and \cite{Mak} and references therein, or axially symmetric, see e.g.\cite{Berch2}, \cite{Berch1}, \cite{Khan} and \cite{Berch}.
The models with broken axial symmetry usually implied that stellar density is uniform on triaxial elliptical shells, see e.g. \cite{Vas2} and \cite{Vas1}. All these 
symmetries are likely to be absent in a realistic stellar environment of SBBH
\footnote{Let us stress that the symmetries
of the stellar distribution correspond to its initial state, since they are broken to some extent after
a SBBH starts to evolve within a stellar cluster.}. 

In this Paper we would like to analyze the evolution of SBBH binary consisting of a more massive black
hole (a primary) with mass $M_{p}$ surrounded by a nuclear star cluster (NSC) and a less massive intruding black hole (a secondary) with mass $M_{s} \ll M_{p}$. The star cluster is assumed
to have a non-spherical distribution of stellar density at the scales of order of the radius of influence of the primary, $r_{infl}$, which may provide a perturbation to the gravitational potential of the
primary at much smaller scales. Note that, in order to have a non-trivial perturbation of
gravitational potential the distribution of stellar
density shouldn't be uniform on elliptical shells, since it is well known that gravitational potential
inside such a shell is uniform, see e.g. \cite{Chandr}.     Therefore, we are going to make the essential assumption that the stellar distribution deviate significantly from being uniform on the elliptical shells,  although its exact form is not important for our purposes. After this assumption is made the non-spherical part of gravitational potential may cause
a secular orbital evolution at scales smaller than $r_{infl}$. 
In case when the stellar distribution is non-spherical only at scales $\sim r_{infl}$, at much smaller scale the radial dependence
of the perturbing gravitational potential, $V_p$ in the quadrupole (or, tidal) approximation is quadratic, while its angular dependence is proportional to a linear
combination of spherical harmonics of the degree $l=2$ with the azimuthal number $m$ ranging from $-2$ to $2$,
\begin{equation}
V_c\propto r^2\sum_{m=-2}^{2}\alpha_{m}Y_{2,m}(\theta, \phi),
\label{e0}    
\end{equation}
where $\theta$ and $\phi$ are polar and azimuthal angles in spherical coordinate system.
The requirement that the gravitational potential is real links the terms 
with negative and positive $m$ by the requirement that $\alpha_{-m}=(-1)^{m}{\alpha_{m}}^{*}$, where $(^*)$ stands for complex conjugate.
Therefore, the angular dependence can be characterized by only positive values of $m$. Moreover, by a rotation of coordinate system
the terms with $m=1$ can be eliminated, see e.g. \cite{IP} for the corresponding expressions. Thus, a general non-uniform distribution
of stellar density leads to the presence of terms proportional to the spherical harmonics
with $m=0$ and $m=2$ as well as their complex conjugate
\footnote{Note that when density perturbations are small, the explicit form of
$\alpha_m$ can be found e.g. in \cite{BT}}.

Thus, we expect that at scales
$\ll r_{infl}$ the orbital dynamics is mainly governed by the Newtonian potential of the
primary and the influence of the non-spherical distribution of stars as well as other
factors (e.g. the relativistic corrections to the Newtonian potential, gravitational potential of stars
within the orbit, etc.). This influence can be treated
as perturbing corrections that cause a secular evolution of the orbit. On the other hand, dynamical friction is expected to be efficient only when a typical stellar mass enclosed within the binary orbit,
$M_{st}$, is larger than the mass of the secondary black hole,
and 
becomes inefficient at scales smaller 
than the the scale $r_{df}$ corresponding to the condition $M_{st}(r_{df})=M_{s}$, see 
\cite{IPS} and \cite{Mak}. In our case $M_{s}\ll M_{p}$, therefore,
the scale $r_{df}$  should be smaller than $r_{infl}$, and therefore the secular evolution appears to be important precisely at the scales, where the orbital evolution due to dynamical friction is expected
to stall. 

The term in the perturbing potential determined by the azimuthal number $m=0$ is axially symmetric. Its presence leads to the secular evolution of the ZLK type. This evolution has been studied
by many authors (see, e.g. \cite{IPS} for a simple approach in a similar setting). Since
the axial symmetry implies that the projection of orbital angular momentum onto the axis
perpendicular to the symmetry plane is conserved, the secular evolution of eccentricity
due to this term is limited by this conservation law. On the other hand, the secular evolution that arises from the presence of the term determined by $m=2$ does not conserve
any component of angular momentum. Its presence can lead, in principle, to a secular 
evolution, which can bring initially small or moderate eccentricity to values close
to unity. A binary with a highly eccentric orbit can evolve very fast due to the emission of gravitational waves. The purpose of this paper is to 
consider such a possibility using analytic and semi-analytic methods. We also will give
some order of magnitude estimates, which may be appropriate for some interesting astrophysical objects.

To the best of our knowledge, such a problem has not been considered in the context of secular orbital evolution of SBBHs However, it has been considered by several authors
in the context of secular evolution of stellar mass binaries orbiting in a star cluster, see e.g \cite{Petr}, \cite{Hami1}, \cite{Hami2}
and \cite{Hami3}. To formulate Hamiltonian describing the secular dynamics of orbital elements
due the presence of non-spherical stellar distribution we use the averaged over the mean anomaly expression for the perturbing potential
obtained in \cite{Hami1}. 

Since, in general, equations describing the secular evolution determined by the $m=2$ terms cannot
be integrated analytically even in the simplest case when the primary gravitational potential is 
Newtonian and all other perturbing factors are neglected, we are going to make a number of rather
drastic assumptions to make our problem analytically treatable. At first, we formally assume that all stars in the vicinity of the secondary are quickly dispersed and neglect their gravitational field. Secondly, we mainly concentrate on the dynamics determined by the $m=2$ terms and only briefly discuss effects arising from the additional inclusion of the $m=0$ term. Thirdly, we formally assume in our analytic study that there is a stage in the evolution of the dynamical system when both eccentricity $e_0$ and inclination to the plane of symmetry $i_0$ are small. This assumption allows us to find some approximate analytic solutions. Fourthly, we neglect all relativistic corrections apart from the effect of Einstein precession of the apsidal line, which is treated as a perturbation leading to a slow evolution of the parameters describing our approximate analytic solutions obtained for the purely Newtonian problem.

We find that the system exhibit quasi-periodic behavior with quasi-periodic cycles with periods
order of the characteristic time of the ZLK effect. In the limit $i_0 \rightarrow 0$, during half of the period the inclination $i$ is close to zero and it is close to $\pi$ during another half. These stages
are separated by relatively short time intervals of growth or decrease of the inclination. These time intervals tend to zero when $i_0 \rightarrow 0$.  
Twice per cycle the value of inclination passes through  $\pi/2$, the first time when the inclination increases from $0$ to $\pi$, and the second time on its way back. 
The eccentricity $e$ also reaches its maximal values twice per cycle, at times close to the times corresponding to inclination being equal to $\pi/2$. The maximal values are inversely proportional to $i_0$.  They also turn out to be functions of $\omega_0$ - the nodal angle at the moment of time when 
inclination and eccentricity have their minimal values per cycle, $e=e_{0}$ and $i=i_{0}$.

Corrections resulting from next order terms in the expansion over $i_0$ and $e_0$ in the equations of motions as well as from other factors such as the Einstein precession lead to a slow change of $i_0$, $\omega_0$ and $e_0$. As a results of it the maximal values of $e$ can be very close to unity at a time much larger than the cycle period (say, one hundred times larger, see below). In case when the slow evolution is determined by the Einstein precession, arguments based on the conservation of full Hamiltonian  of the problem lead to an estimate of the maximal possible value of eccentricity, which turns out to the proportional to the rate of Einstein precession.

The analytic results provide some  important qualitative understanding of the general dynamics of the system, but, they are directly applicable only when $i_0$ is quite small. Since a number of
binaries with a small $i_0$ is expected to be quite small, the direct application of the analytic
results of the problems of a rapid circularization due to emission 
of gravitational waves (hereafter, GWs) gives a small number of such events. We show, however, using numerical means, that systems with rather large initial inclinations can also experience a similar dynamics. This is partly explained by the observation that even in this case when the system evolves during
a sufficiently long time there are extended periods, which correspond to the cycles with rather small minimal inclinations and, accordingly, rather large maximal values of eccentricity.

These results are used to estimate the possibility of an efficient circularization due to emission of GWs.
The efficient circularization is defined as having the characteristic circularization timescale being order of or smaller than the orbital period. In our estimates we use an empirical 
relationship between NSC size and a mass of the central black hole reported in \cite{Georg} for 
early type galaxies under the assumption that $r_{infl}$ has the same order of magnitude as
the NSC size, see also \cite{Georg}. We also assume that the stellar density distribution inside NSC has 
the standard Young profile \cite{Young}. We show that in our model the efficient circularization is expected at times order of or smaller than a few Gyrs provided that the mass ratio of SBBH, $q$, is
sufficiently large, $q > \sim 0.04$. It is important to point out that, for
certain parameters of the problem, orbital periastron in the beginning of the circularization phase is estimated to be as small as a few gravitational radii. That means that,  in this case, the orbital eccentricity remains to be relatively larger all
the way down to the black holes merger. This would allow one to distinguish GW events originating
from our mechanism from the ones related to the standard hardening process by a different shape of
GW signal. This also may be relevant for the formation of binaries with the parameters similar to
the ones assumed in the standard model of OJ 287. 

The plan of the Paper is as follows 
1) In Section 2
we introduce equations of motion describing the secular evolution of SBBHs orbit 
 due to the action of the perturbing potential and the Einstein apsidal precession.

2) In Section 3
we provide the asymptotic solution of the equations of motion in the purely
Newtonian 
case, taking into account only the $m=2$ terms in the perturbing potential. We also provide
some brief numerical analysis of the role of the $m=0$ term and argue that it shouldn't change qualitatively the behavior of the system if 
its contribution is relatively small in Appendix \ref{phi}. 

3)In Section 4 
we analyze the role of Einstein precession  
and estimate  
the maximal value of eccentricity, which could be reached in the course of SBBH evolution.
It is shown that its role can be described as a map between parameters of consequent cycles
of the Newtonian secular evolution and a technical analysis of this map is made in Appendix
\ref{map}.
We also provide some numerical analysis, which is
used to justify our assumption that our analytic results  
are applicable to 
sufficiently large inclinations $i_0$ when a sufficiently 
large enough evolution time is considered.

4) In Section 5
we 
estimate   
parameters of SBBHs and NSCs required for 
efficient orbital circularization due to GWs emission.

We conclude that relatively large mass ratios $q >\sim 0.04$ are preferable. We also provide an upper estimate of a typical time elapsed before such circularization takes place. 

5)
We summarize and finally discuss our results in 
Section 6.

Note that a reader who is interested in astrophysical applications only can omit Section 2-4 and go directly to Section 5, which is written in a self-consistent way.

\bigskip
\section{The secular evolution of SBBHs 
in the presence of quadrupole perturbing potential}

\bigskip
Let us consider 
the orbit of 
the secondary black hole with mass $M_s$
in gravitational potential $V=V_p+V_c$, where $V_p$ is the gravitational potential of 
the primary black hole with mass $M_p$ and $V_c\ll V_p$ is a perturbing potential generated by the outer parts of the stellar cluster with a non-symmetric
distribution of stellar mass density. Let us start for simplicity from Newtonian approximation, when $V_p =GM_p/r$ (general relativistic corrections will be considered below). As have already been discussed in Introduction we assume in this Paper that 
$V_c$ has the quadrupole form. Under this assumption in Cartesian
coordinate system $(x,y,z)$, $V_c$ of the form (\ref{e0}) can be represented as:
\begin{equation}
V_c=\Omega^2[\sin\nu (2z^2-(x^2+y^2))+\cos\nu (x^2-y^2)],
\label{q1}
\end{equation}  
where the constant $\nu$ determines the
relative contribution of the two terms in  
square brackets, $\Omega$ has the dimension of inverse time, 
it is assumed to be smaller than mean motion 
\begin{equation}
n_0=\Omega \ll\sqrt{GM_p\over a^3},
\label{n0}
\end{equation}
where $a$ is the semi-major axis of the secondary black hole orbit. 
The first term in square brackets in (\ref{q1}) is determined by the harmonics of the potential with azimuthal number $m=0$. It has, accordingly, the azimuthal symmetry, which conserves the projection of orbital angular momentum onto the axis $z$. It can be shown (see e.g. \cite{Hami1}), that the secular evolution due to the presence of this term has the standard 
ZLK character. The second term is due to non-axially
symmetric part of the potential, which is determined by harmonics with $m=\pm 2$. It leads to a non-standard secular evolution, 
which does not conserve any component of orbital angular momentum. 

We parametrize the orbit by the standard set of variables including orbital semi-major axis $a$, eccentricity $e$, inclination
$i$, mean motion $n_0$,
argument of pericentre $\varpi$ and longitude of ascending node $\omega$
\footnote{We remind that $n_0$ is also often denoted as the orbital angular frequency, 
$i$ is defined as the inclination angle of the orbital
plane with respect to the symmetry plane of the problem, $\omega$ defines the position
of the intersection line of the orbital and symmetry planes in the symmetry plane,
and $\varpi$ defines the position of the orbital periastron with respect 
to the intersection line, see e.g. 
https://en.wikipedia.org/wiki/Orbital$\_$node$\#$Node$\_$distinction for a graphical representation.}. In order to bring equations
of motion to the canonical form one can use 
Delaunay variables (see for example \cite{Carl}) 
 \begin{equation}
P_1=\sqrt{GMa}, \quad P_2=\sqrt{(1-e^2)}P_1, \quad P_3=\cos i P_2,
\label{q2}
\end{equation}  
which are generalized momenta canonically conjugate to $n_0$, $\varpi$ and $\omega$, respectively.

We use the average of (\ref{q1}) over 
the mean anomaly $n_0t$ using equations (A1-A3) from \cite{Hami1}, the resulting expression is
\begin{equation}
H=\Omega^2[\sin\nu (2<z>^2-(<x>^2+<y>^2)+\cos\nu (<x>^2-<y>^2)],
\label{q3}
\end{equation}
where $<..>$ stand for the averaged quantities, provides Hamiltonian for the secular motion. 

The corresponding equations
of motion follow from Hamilton equations and have the form
\begin{align}
{de\over dt}={\epsilon\over P_1e}{\partial \over \partial \varpi}H, \quad {di\over dt}={1\over P_1\epsilon \sin i}({\partial \over \partial \omega}-\cos i {\partial \over \partial \varpi})H, 
\nonumber \\
{d\varpi \over dt}=-{1\over P_1\epsilon}(\cos i {\partial \over \partial \cos i}+{\epsilon^2\over e} 
{\partial \over \partial e})H, \quad
\quad {d\omega \over dt}={1\over P_1\epsilon}  {\partial \over \partial \cos i}H, 
\label{q5}
\end{align}
where $\epsilon=\sqrt{1-e^2}$ and we note that, since the problem is stationary,
the orbital energy and, accordingly, 
semi-major axis $a$ and $P_1$ are conserved.
It is convenient
to divide the full 
Hamiltonian $H$ into two parts $H=\sin\nu H_0+\cos\nu H_2$, where $H_0$ and $H_2$ are proportional to
$2<z>^2-(<x>^2+<y>^2)$ and $<x>^2-<y>^2$), respectively.
We also represent time derivatives of all our dynamical variables as the sums 
\begin{equation}
{dq\over dt}=\nu_o ({dq\over dt})_0+\nu_2 ({dq\over dt})_2,
\label{q4}
\end{equation}
where $q$ stands for any of $e$, $i$, $\omega$ and $\varpi$,   $({dq\over dt})_0$ and $({dq\over dt})_2$ are given by eqns (\ref{q5}) when $H$ in these equations  is replaced by $H_0$ and $H_2$, respectively.

From the results reported in \cite{Hami1}
we obtain
\begin{equation}
H_0={\Omega^2a^2\over 4}h_0, \quad h_0=((2+3e^2)(1-3\cos^2 i)-15e^2\sin^2 i\cos 2\varpi),
\label{q6}
\end{equation}
and 
\begin{align}
H_2={\Omega^2a^2\over 4}h_2, \quad h_2=(\cos^2i\cos 2\omega (5e^2\cos 2\varpi -(2+3e^2)) \nonumber \\
-10e^2\cos i\sin 2\omega \sin 2\varpi 
\nonumber \\ +
\cos 2\omega (5e^2\cos 2\varpi + 2+3e^2)). 
\label{q7}
\end{align}

We obtain from eq. (\ref{q6})
\begin{align}
({de\over d\tau})_0=30\epsilon e\sin^2 i\sin 2\varpi \nonumber \\
({di\over d\tau})_0=-{30e^2\over \epsilon}\sin i\cos i\sin 2\varpi \nonumber \\
({d\varpi \over d\tau})_0=-{6\over \epsilon}(5\cos^2 i(\cos 2\varpi -1)+\epsilon^2 (1-5\cos 2\varpi)) \nonumber \\
({d\omega \over d\tau})_0={6\over \epsilon}(5e^2\cos 2\varpi -(2+3e^2)), 
\label{q8}
\end{align}
where 
\begin{align}
\tau=t/t_{*}, \quad t_{*}={4n_{0}\over \Omega^{2}}, \quad n_{0} ={\sqrt {GM_{p}\over a^{3}}.}
\label{q8n}
\end{align}
Eqns (\ref{q8}) describe the standard ZLK
evolution (see e.g. \cite{IPS}, hereafter IPS). To see this, it is enough to make a simple replacement of variables in (\ref{q8}), namely, replace  $\tau$ with $t=t_*\tau$. 
From eq. (\ref{q7}) we have
\begin{align}
({de\over d\tau})_2=-10\epsilon e (\cos 2\omega \sin 2\varpi (1+\cos^2 i)+2\cos i \sin 2\omega \cos 2\varpi)     
\label{q9a} \\
({di\over d\tau})_2= -{2 \sin  i\over \epsilon}(5e^2\cos i\cos 2\omega \sin 2\varpi +\sin 2\omega (5e^2\cos 2\varpi +2 +3e^2))    
\label{q9b} \\
({d\varpi \over d\tau})_2= -{2 \over \epsilon}(\cos 2\omega (5\cos 2\varpi (\epsilon^2+\cos^2 i)+3\epsilon^2-5\cos^2 i)
-5(2-e^2)\cos i\sin 2\omega \sin 2\varpi) \label{q9c} \\
({d\omega \over d\tau})_2 = {2 \over \epsilon}(\cos i \cos 2\omega (5e^2\cos 2\varpi -(2+3e^2))-5e^2\sin 2\omega \sin 2\varpi). 
\label{q9d}
\end{align}

Since we consider a binary black hole, we need to take into 
account relativistic corrections to the Newtonian motion. Since the orbital semi-major axis is much larger than the gravitational radius of the central (i.e. primary) black hole, it is enough for our purposes to take into account the first-order correction, which is the Einstein precession of the apsidal angle. This 
rate is given by the standard expression, see for example \cite{LL} 
\begin{align}
{\dot \varpi_{E}}
 ={3GM\over c^2a(1-e^2)}n_{0}. \label{Ein}
\end{align}
where $c$ is the speed of light.
In terms of $\tau$ (see 
eq. (\ref{q8n}) eq. (\ref{Ein}) can be rewritten as
\begin{align}
{d\varpi_{E}\over d\tau}
 =\beta_{E}\epsilon^{-2}, \quad \beta_{E}={12GM\over c^2 a}{n_0^2\over \Omega^2}
\label{Ein1},
\end{align}
 where $\epsilon$ is defined after eq.(\ref{q5}).
It is possible to see that the corresponding contribution to Hamiltonian of our system has the form
\begin{align}
H_{E}={\Omega^2a^2\over 4}h_{E}, \quad h_{E}=-{\beta_{E}\over \epsilon}.
\label{Ein2}
\end{align}

Note that, in principle, it is easy to incorporate in our dynamical system the next order relativistic correction due
to the Lense-Thirring precession of the nodal angle. The corresponding precession rate of this angle, $\dot \omega_{LT}$,  is proportional
to the primary black hole rotational parameter $\chi$ (such that $|\chi| <1$). Its explicit expression can be found in e.g. 
\cite{Mer13}:
\begin{align}
\dot \omega_{LT}={2\chi G^2 M^2 \over c^3 \epsilon^3 a^{3}}.
\label{LT1}
\end{align}
From the condition $|\dot \omega_{LT}| < \dot \varpi_E$ we readily find
\begin{align}
{GM (1+e)\over c^2 r_p} < {9\over 4}{1\over |\chi|^2},
\label{LT2}
\end{align}
where $r_p=(1-e)a$ is the orbital periastron. It is shown below that our dynamical system experiences secular changes of 
eccentricity between a state, where $e$ is small and ${GM\over c^2 r_p} \ll 1$ and a state, where $e$ is close to unity. 
But, even for the latter state ${GM\over c^2 r_p} $ is expected to be smaller than ${1\over 4}$, see eq. (\ref{q66}) below. Thus,
the inequality (\ref{LT2}) is satisfied for all potentially interesting states of the system, and, therefore, for simplicity,
we neglect the evolution of the nodal angle caused to the Lense-Thirring effect below.

Accordingly, the full Hamiltonian of our dynamical system, $H$, can be represented as
\begin{align}
H={\Omega^2a^2\over 4}h, \quad h=(\sin (\nu) h_0+\cos (\nu) h_2) + h_{E}.
\label{Htot}
\end{align}

The full equations of motion following from (\ref{Htot}) have symbolic form analogous to (\ref{q4}) 
\begin{equation}
{dq\over d\tau}=\sin \nu ({dq\over d\tau})_0+\cos \nu ({dq\over d\tau})_2+({dq\over d\tau})_{E},
\label{eqtot}
\end{equation}
where by $({dq\over d\tau})_{E}$ we imply that (\ref{Ein1}) should be added to the equation describing the
rate of apsidal precession.

\bigskip

\section{Numerical and asymptotic analytic solutions of the secular equations in the Newtonian approximation}

\bigskip
Since in this Paper we would like to discuss only a qualitative character of the evolution of the binary under the influence
of the perturbing potential and a principal possibility of the gravitational capture of the secondary BH on close-in
orbits, we are going to make a number of rather drastic simplifying assumptions. In particular, 
in this Section we consider a simplified model problem, where the potential is determined by only $\pm 2$ terms in equation (\ref{q1}) and the Einstein precession is negligible. 
We set, accordingly, $\nu=0$
and  $({dq\over d\tau})_{E}={d\varpi_E\over d\tau}=0$ in eq. (\ref{eqtot})
However, 
 we briefly discuss effects
originating from  $\nu \ne 0$ in Appendix \ref{phi}. 
In this case the secular evolution
is determined 
by equations (\ref{q9a}-\ref{q9d}).

\bigskip
\subsection{The asymptotic numerical solution  in the limit $i_{0}\rightarrow 0$}
\begin{figure}
    \includegraphics[width=0.5\linewidth]{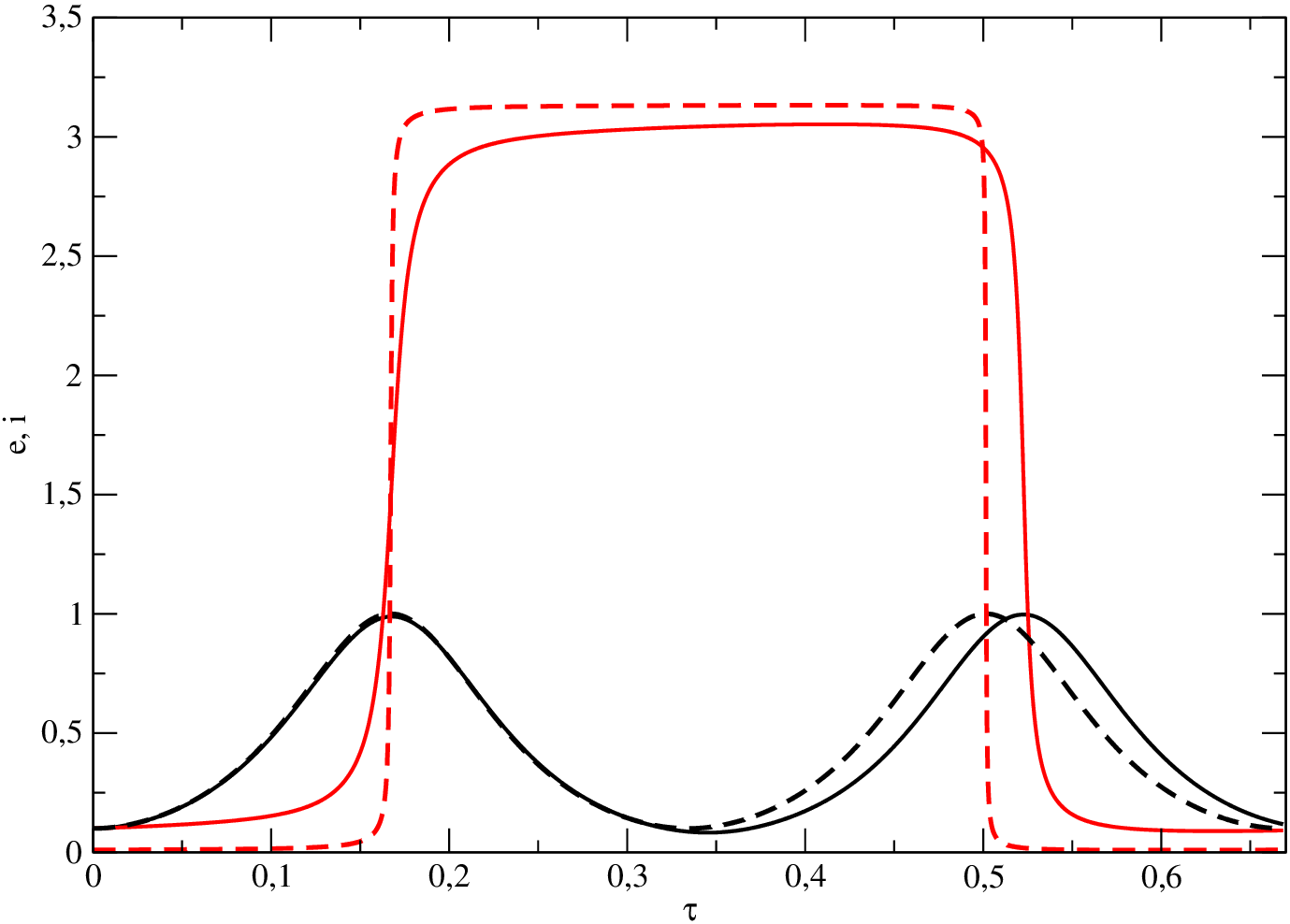}
    \includegraphics[width=0.5\linewidth]{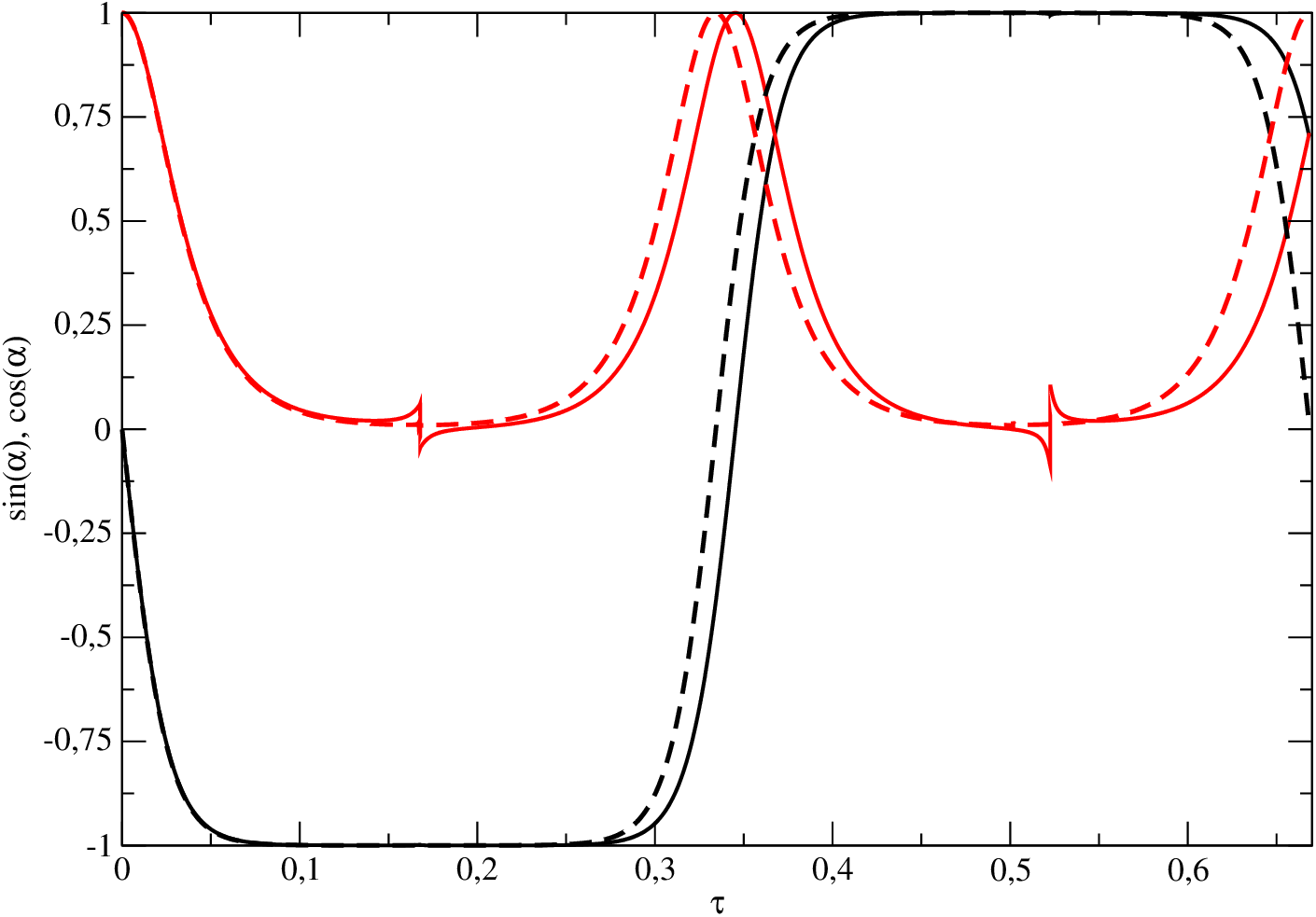}
\caption{Top panel. The numerical dependencies of $e$ (the black curves) and $i$ (the red curves) as functions of time, the initial value of eccentricity, apsidal 
and nodal angles, $e_0=0.1$, $\varpi_0=2\pi/5$ and $\omega_0=-\varpi_0=-2\pi/5$, respectively. Note that the particular values of $\varpi_0$ and $\omega_0$ are chosen
for illustrative purpose only, we checked that other values result in a similar evolution.
Also note that the condition $\alpha_0=2(\omega_0+\varpi_0)=0$ ensures that the initial value of eccentricity is the minimal one in the course of evolution.
Solid and dashed curves correspond to initial inclination $i_0=0.1$ and $0.01$, respectively. Bottom panel. The evolution of 
$\sin (\alpha)$ (the black curves) and $\cos (\alpha )$ (the red curves), the initial values of the state variables and types of the curves are the same as in top panel.}
\label{Fig1}
\end{figure}

A typical example of the numerical solution is illustrated in Fig. \ref{Fig1}, where we show 
the solution corresponding to small initial eccentricity $e_0=0.1$, and two small values of initial inclination $i=0.01$ (solid lines) and $i=0.1$ (dashed lines), respectively. The top panel shows the evolution of $e$ and $i$ 
as the black and red curves, respectively, while the bottom panel describes the evolution of $\omega$ and $\varpi$.
For this purpose, we find it convenient to introduce a new angle, $\alpha$, defined according to the rule: $\alpha=2(\omega+\varpi)$ when
$i < \pi/2$ and $\alpha=2(\omega-\varpi)$ when $i > \pi/2$. We show the evolution of $\sin (\alpha)$ as the black curves 
and  $\cos (\alpha)$ as the red curves.

One can see from Fig. \ref{Fig1} that the evolution of $e$ is almost periodical with the eccentricity gradually increasing from
the initial small value to a value $e\approx 1$, then decreasing back to approximately the same minimal value; then this cycle
repeats. On the other hand, $i$ during an extended initial period of time has small values, then it sharply grows to a value approximately equal to $\pi $, then it stays approximately constant again, then it sharply
drops to small values and approaches values
close to $i_0$ at the end of the calculation. Thus, we have approximately two periods of evolution of $e$ and one period of evolution of $i$ during computational time. It is very important to note that the fast time evolution
of $i$ from $\sim 0$ to $\sim \pi$ and back occurs during time periods corresponding to $e\sim 1$, with its typical time duration decreasing with decreasing $i_0$. Apart from these time periods the curves corresponding to the different values of $i_0$ are
close to each other. The curves showing the evolution of $\sin (\alpha)$ and $\cos (\alpha)$ are also quasi-periodic. Also, apart
from  a small 'kink-like' feature seen in the curve describing the 
evolution of $\cos (\alpha)$ when it is
close to zero these quantities smoothly transit through the time
periods of the fast evolution of $i$. We have $\sin (\alpha)\approx -1$ and $\cos (\alpha)\approx 0$ when $i$ transits from
$\sim 0$ to $\sim \pi$ and  $\sin (\alpha)\approx 1$ and $\cos (\alpha)\approx 0$  when $i$ transits from
$\sim \pi$ to $\sim 0$. Note that  $\sin (\alpha)$ changes its sign approximately in the middle of the shown period of time,
when $i\approx \pi$ and $e\approx e_0$.

These numerical results suggests the following analytical scheme for a treatment of the case with small initial
values of eccentricity and inclination. At first, we consider the limit $i_0 \rightarrow 0$ \footnote{Note that it is qualitatively
different from the case $i_0=0$, where $i$ remains to be zero during the evolution, see eq. (12).}. In this case the
numerical results indicate that the evolution can be considered as consisting of subsequent stages with $i\approx 0$ and $i\approx \pi$,
respectively, with sharp jumps of $i$ between them and 
the duration of these jumps
tends to zero when  $i_0 \rightarrow 0$. 
In contrast, in the same limit $i_0\rightarrow 0$ 
$\alpha$ is nearly 
a constant during the fast evolution of $i$.  
This behavior of $i$ can be used to construct an analytic asymptotic solution to equations (\ref{q9a}) and (\ref{q9c}-\ref{q9d}).


\subsection{The asymptotic analytical solution in the limit $i_{0}\rightarrow 0$}

It is then turns out to be 
possible to consider a finite, but small $i_0$, and describe analytically  the evolution within the region of sharp jumps
using the simplifying assumptions that $e\sim 1$, $\cos (\alpha)\approx 0$ and $\sin (\alpha) \approx \mp 1$ during these special
periods of time. 
Here we would like to consider a model problem, which provides a zero-order approximation to the asymptotic solution of our dynamical equations
in the limit  $i_{0}\rightarrow 0$. We start with some initial values of $e_0 \ll 1$, $\omega_0$ and $\varpi_0$ and assume that the inclination angle $i=0$ until we reach the state with $e=1$ at some time $\tau=\tau_*$, which is expected to be finite.    
We then set $i=\pi$, $e=1$ and assume that the angle $\alpha$ is the same as it was in the end of the {p{preceding}} 
stage. The
eccentricity is expected to decrease to some minimal value, $e_{min}$, and then increase again to the state with $e=1$. The inclination
angle is again switched to $i=0$ and the cycle is ended when $e$ reaches the value $e_0$. 
It follows from our previous discussion 
see equation (\ref{q12}), that
in order to have $e_{min}=e_{0}$ we should have $\alpha=0$ and, accordingly, $\varpi_0=-\omega_0$.
Thus, this model problem is characterized by two parameters, 
$e_0$ and $\omega_0$.

Setting $i=0$ and $\pi$ we have $\cos(i)=\pm 1$. It then follows from equations (\ref{q9a}) and (\ref{q9c}-\ref{q9d}) that
\begin{align}
{de\over d\tau}=-20\epsilon e \sin 2 (\varpi \pm \omega),
\label{q10a} \\
{d(\varpi \pm \omega)\over d\tau}=-20\epsilon  \cos 2 (\varpi \pm \omega).    
\label{q10b} 
\end{align}
Additionally, from eq. (\ref{q9d}) we obtain  
\begin{align}
{d\omega \over d\tau}=\pm {2\over \epsilon}(5e^2 \cos 2 (\varpi \pm \omega)-(2+3e^2)\cos 2\omega).
\label{q11} 
\end{align}
It is easy to see 
that eqns (\ref{q10a}-\ref{q10b}) can be also rewritten as
\begin{align}
{de\over d\tau}=\mp 20\epsilon e \sin \alpha, {d\alpha \over d\tau}=\mp 40\epsilon  \cos \alpha.    
\label{q10n} 
\end{align}
Dividing (\ref{q10a}) by (\ref{q10b}) (or, the first eq. in (\ref{q10n}) by the second one)  
we see that the quantity $C=e^2|\cos  (\alpha)|$ is conserved. Note that the same conservation law follows from 
(\ref{q7}) when $i=0$ or  
$i=\pi$.

Since equations (\ref{q10a}-\ref{q10b}) form a complete set we can first solve them in all regions separately and then 
match all these solutions 
together keeping in mind that   
$e$ and $\alpha$ should be continuous during sharp jumps mentioned above. 
After that we can 
solve equation (\ref{q11}). 

We assume that when $\tau=0$ the  eccentricity $e$ is in its local minimum, $e=e_{0}$.
Hence, as it follows from eq. (\ref{q10n}) 
that $\alpha=0$ or $\pi$ when $\tau=0$, and, accordingly, $C=e_{0}^2$.
It also follows from eqns (\ref{q10n})
that they
are invariant with respect to the change $\alpha \rightarrow \alpha + \pi$, $\tau \rightarrow -\tau$, and, therefore, the case $\alpha(\tau)=0$
and $\alpha(\tau=0)=\pi$ can be obtained from each other by the change
of the sign of time.
Therefore, we consider below only the case $\alpha(\tau=0)=0$. In
this case we expect 
that $\cos  (\alpha) > 0$ in the course of the whole evolution and get $e^2_0=e^2\cos  (\alpha)$. 
Using this relation we can express $\cos (\alpha )$ and $\sin (\alpha )$ in terms
of $e$ as
\begin{align}
\cos \alpha = ({e_0\over e})^2, \quad \sin \alpha =\kappa \sqrt {1-{({e_0\over e})}^4},
\label{q12} 
\end{align} 
where $\kappa $ is equal to $\pm 1$ and the sign should be chosen from the following arguments. Initially, eccentricity should grow
with time
from $e_0$ to $1$. We also have $i=0$ initially, and, therefore, we should choose the sign $(+)$ in eq. (\ref{q10a}). Thus, at the initial stage $\kappa$ should be equal to $-1$. After the state $e=1$ is reached we must choose the sign $(-)$ in eq. (\ref{q10a}). Since
$\alpha$ is assumed to pass continuously through the transition from one stage to another, $\kappa $ remains negative and the eccentricity,
accordingly, decreases
at times $\tau > \tau_{*}$. This happens until the state with $e=e_0$ is reached and the eccentricity must increase after the corresponding time. Therefore, we must change the sign of $\kappa$ and set $\kappa=1$ when $e=e_0$ at the stage with $i=\pi$\footnote{Let us stress that the choice to $\kappa$ to be equal to either (-) or (+) is { independent} from
the choice of $\cos (i)=\pm 1$ during the stages when $i$ is either close to zero or $\pi$. In the former case
we use $\kappa$, while in the latter case we show explicitly $(\pm)$ in all appropriate equations.}. 
It is then follows from eq. (\ref{q10a}) that the eccentricity increases again up to the state with $e=1$, we then change again
$(-)$ to $(+)$ in eq. (\ref{q10a}) after the corresponding time, while keeping $\kappa=1$. We have again the stage with $i=0$ after this moment of time, but with decreasing eccentricity. The eccentricity decreases down to $e=e_0$ and after the corresponding moment of time the cycle repeats.

Substituting eq. (\ref{q12}) in eq. (\ref{q10a}) we formally obtain
\begin{align}
\int^{e}_{e_0} {de \over  e\epsilon \sqrt {1-{({e_0\over e})}^4}}=\mp 20\kappa \tau
\label{q13} 
\end{align} 
Evaluation of the  integral on l.h.s leads to a rather cumbersome expression. However, for our purposes it would enough to evaluate
it only in the limit $e \gg e_0$. Let us consider, for definiteness, the initial stage with $i=1$ and $\kappa=-1$, where the
expression on r.h.s. of (\ref{q13}) is positive. In this case we can divide the range of integration on two parts $e_{0} \le e \le e_{*}$ and $e_{*} \le e \le 1$, where an intermediate value of $e_*$ is chosen in such a way that $e_0 \ll e_{*} \ll 1$. When considering the integral
in the range  $e_{0} \le e \le e_{*}$ we can set $\epsilon=1$, while in the range  $e_{*} \le e \le 1$ we can set
$\sqrt {1-{({e_0\over e})}^4}=1$ in the integrand. After these approximations are made the corresponding integrals are elementary. Moreover, their sum
does not depend on $e_{*}$ in the leading order over $e_0/e_{*}$. In this way we obtain
\begin{align}
\epsilon={(1-{e_0^2\over 8}\exp (40\tau))\over (1+{e_0^2\over 8}\exp (40\tau))},
\label{q14} 
\end{align} 
and, from the condition $e(\tau_{*})=1$ we find that
\begin{align}
\tau_{*}={1\over 40}\log {8\over e_{0}^2}.
\label{q15} 
\end{align}
The time $\tau_*$ is the characteristic time of the process under consideration and we consider in this Section
the evolution of our system at times comparable to $\tau_{*}$. 

When $\tau > \tau_*$ we should choose $(-)$ on r.h.s. of eq. (\ref{q13}) and specify the constant of integration in such a way
that $e(\tau=\tau_{*})=1$. This branch of the solution corresponds to decreasing with time eccentricity and can be obtained
from eq. (\ref{q14}) by changing $\tau$ to $2\tau_*-\tau$ due to symmetry arguments. It is clear that $e=e_0$ when $\tau=2\tau_*$.
At this time we should change $\kappa$ from $-1$ to $1$, and at times $\tau > 2\tau_{*}$ the solution possesses again growing
eccentricity and can be obtained from eq. (\ref{q13}) by the substitution $\tau-2\tau_{*}$ instead of $\tau$. This branch describes
the evolution of eccentricity until $\tau=3\tau_*$ when we have $e=1$ again. At this time we should again switch to the solution
with decreasing eccentricity, which can be obtained in the same way as before, but changing $\tau$ in eq. (\ref{q13}) to  $4\tau_*-\tau$.
The latter branch describes the solution until $\tau=4\tau_{*}$. 
It is clear that the evolution of eccentricity is periodic with the period $2\tau_{*}$, but
$\sin (\alpha)$ in eq. (\ref{q12}) is negative when $\tau < 2\tau_{*}$ (or, $\kappa=-1$) and positive otherwise
($\kappa=1$). Thus, the whole period of the evolution
of all state variables in the considered limit is $4\tau_{*}$. Using these rules we obtain explicitly
\begin{align}
\epsilon={(1-{e_0^2\over 8}\exp (40\tau))\over (1+{e_0^2\over 8}\exp (40\tau))} \quad when \quad \tau < \tau_{*},
\nonumber \\
\epsilon={({e_0^2\over 8}\exp (40\tau)-1)\over (1+{e_0^2\over 8}\exp (40\tau))} \quad when \quad \tau_{*} < \tau < 2\tau_{*},    
\nonumber \\ 
\epsilon={(1-({e_0^2\over 8})^{3}\exp (40\tau))\over (1+({e_0^2\over 8})^{3}\exp (40\tau))} \quad when \quad 2\tau_{*} < \tau < 3\tau_{*},
\nonumber \\ 
\epsilon={(({e_0^2\over 8})^{3}\exp (40\tau)-1)\over (1+({e_0^2\over 8})^{3}\exp (40\tau))}  \quad when \quad 3\tau_{*} < \tau < 4\tau_{*}.
\label{q16}  
\end{align}
Eqns (\ref{q16}) together with eq. (\ref{q12}), where we choose $\kappa = -1$ when $\tau < 2\tau_{*}$ and  $\kappa = 1$ otherwise
describe the whole cycle of the evolution of $e$ and $\alpha$ in the limit of an infinitesimally small initial inclination angle.

In order to describe the separate evolution of $\omega$ and $\varpi$ we should also solve eq. (\ref{q11}). Taking into account
that the expression for $\cos (\alpha)$ in eq. (\ref{q12}) tells that the first term in the brackets in eq. (\ref{q11}) is equal to $5e_0^2$,
which is small everywhere this term can be neglected. We then obtain from eq. (\ref{q11})
\begin{align}
\int_{\omega_{0}}^{\omega}{d\omega \over \cos (2\omega )} ={1\over 10 \kappa}\int^{e}_{e_{0}}{de (2+3e^2)\over e\epsilon^2  (1-{({e_0\over e})}^4)},
\label{q17} 
\end{align}
where we use eq. (\ref{q10a}) to change the integration variable from $\tau$ to $e$. We evaluate the integral on r.h.s. of eq. (\ref{q17})
in the same way as was done in the case of a similar integral in eq. (\ref{q13}). Note that in the case of  eq. (\ref{q17}) there are 
only two branches corresponding to $\kappa=\mp 1$, which should be continuously joined together at $\tau=2\tau_{*}$. In this way
we obtain
\begin{align}
\sin (2\omega)={C^{-}_{\omega}\epsilon^2 -e^{4/5} \over C^{-}_{\omega}\epsilon^2 + e^{4/5}} \quad when \quad \tau < 2\tau_{*},
\nonumber \\
\sin (2\omega)={C^{+}_{\omega}\epsilon^{-2} -e^{-4/5} \over C^{+}_{\omega}\epsilon^{-2} + e^{-4/5}} \quad when \quad 2\tau_{*} < \tau < 4\tau_{*},
\label{q18}  
\end{align}
where 
\begin{align}
C^{-}_{\omega}={(1+\sin (2\omega_{0}))\over (1-\sin (2\omega_{0}))}e_0^{4/5}/(1-e_0^2), \quad C^{+}_{\omega}={(1+\sin (2\omega_{0}))\over (1-\sin (2\omega_{0}))}e_0^{-4/5}(1-e_0^2),
\label{q18p}  
\end{align}
we use $C^{-}_{\omega}$ when  $\tau <2\tau_{*}$ and
$C^{+}_{\omega}$ is used otherwise. 
Sometimes it is convenient
to redefine the angle $\omega$ according to the rule 
\begin{align}
\psi=\omega-{\pi\over 4},
\label{q18a}  
\end{align}  
and rewrite $C^{\kappa}$ as $C^{\mp}=e_{0}^{\pm 4/5}\cot^{2} \psi_0$, where $\psi_0=\psi(\tau=0)$. For our purposes we
also need to know the dependency of $\cos (2\omega)$ on $e$, which can be obtained from
eq. (\ref{q18}) and eq. (\ref{q18a})
\begin{align}
\cos (2\omega)=-\sin (2\psi)=\mp {\sin (2\psi_0)\over \sin^{2}(\psi_0)}{e_0^{2/5}e^{2/5}\epsilon \over (C_{\omega}^{-}\epsilon^2 +
e^{4/5})}
\label{q18b}  
\end{align}   
when $\kappa=-1$ and
\begin{align}
\cos (2\omega)=-\sin (2\psi)=\pm {\sin (2\psi_0)\over \sin^{2}(\psi_0)}{e_0^{-2/5}e^{-2/5}\epsilon^{-1}\over (C_{\omega}^{+}\epsilon^{-2} +
e^{-4/5})}
\label{q18c}  
\end{align}
in the opposite case.
Note that the choice of sign agrees with the fact that $\cos (2\omega)$ should change its sign when
$i$ transits from $0$ to $\pi$ and back. However, we stress that this behaviour is expected only in the
Newtonian case and the sign can be unchanged when the Einstein apsidal precession is taken into account, as
we discuss below. From the expressions (\ref{q18b}) and (\ref{q18c}) it follows that in the limit 
$e\rightarrow 1$ ($\epsilon \rightarrow 0$) we have
\begin{align}
|\cos (2\omega)| \approx 2(C^{-}_{\omega})^{1/2}\epsilon, \quad |\cos (2\omega)| \approx 2(C^{+}_{\omega})^{-1/2}\epsilon.
\label{q23}  
\end{align}

\begin{figure}
    \includegraphics[width=0.5\linewidth]{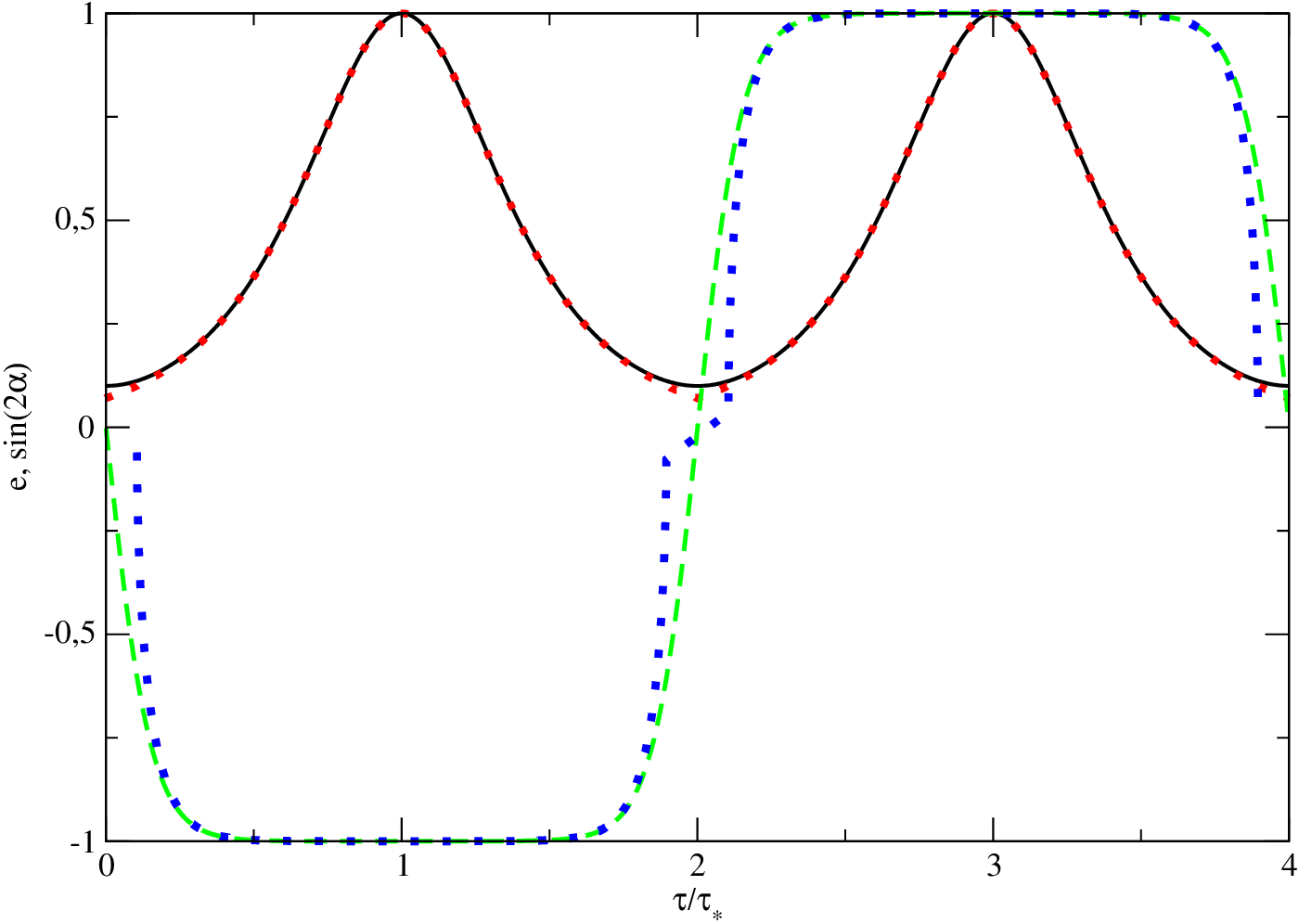}\par 
    \includegraphics[width=0.5\linewidth]{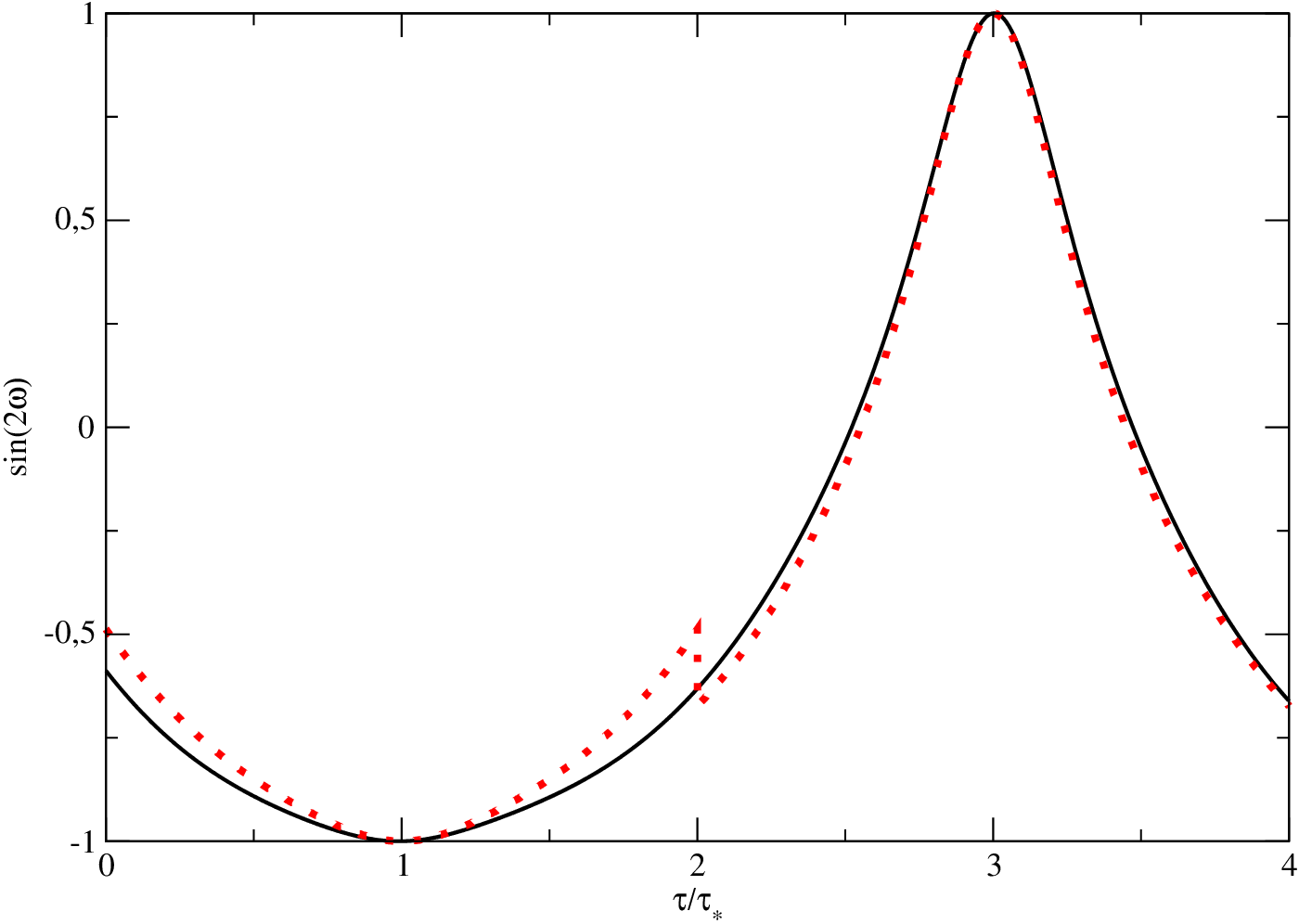}\par 
\caption{Top panel. The numerical dependencies of $e$  (solid line) and $\sin (2\alpha)$ (dashed line) as functions of time, the initial value of eccentricity, inclination
angle, apsidal 
and nodal angles are $e_0=0.1$, $i_0=0.01$, $\varpi_0=2\pi/5$ and $\omega_0=-\varpi_0=-2\pi/5$, respectively. Dotted lines show the analytic expressions
given by eqns (\ref{q12}) and (\ref{q14}). Note that, since it is expected that when $e\sim e_0$ our simple approach  is invalid, the expression under
the square root in (\ref{q12}) can be negative. The corresponding values of time are not shown. 
 Bottom panel. The same as top panel, but for $\sin(2\omega)$. The solid and dotted curves represent the numerical result and the analytic expression
 (\ref{q18}), respectively.}
\label{Fig2}
\end{figure}

We compare the analytic expressions (\ref{q12}), (\ref{q16}) and (\ref{q18}) in Fig. \ref{Fig2}. One can see from this figure that there is
a very good agreement between theoretical and numerical results apart from the regions, where $e\sim e_0$.

\subsubsection{The evolution of inclination in the linear regime}

When the inclination $i$ is either small or close to $\pi$, we can use the solution described above to determine its evolution. For that
we use eq. (\ref{q9b}) setting there  $\cos (i)=\pm 1$, and $\sin (i)=i_{+}$, where $i_{+}\equiv i$ when $i\ll 1$, or $\sin (i)=-i_{-}$, where 
$i_{-}=i-\pi$ when $i\approx \pi$, to obtain
\begin{align}
i_{\pm}^{-1}{di_{\pm}\over d\tau}=-{2i\over \epsilon}(5e^2\sin (2(\varpi \pm \omega )\pm (2+3e^2)\sin (2\omega))={e \over 2\epsilon^2}{de\over d\tau}
\mp {2\over \epsilon}(2+3e^2)\sin (2\omega)),
\label{q19}  
\end{align}
where we use eq. (\ref{q10a}) to obtain the last equality. We use eq. (\ref{q11}) neglecting the first term in the brackets there
to express the last term in (\ref{q19}) as $\tan (2\omega){d\omega \over d\tau}$. After this transformation is made the integration 
of eq. (\ref{q19}) is elementary with the result
\begin{align}
i_{\pm}=C_{i}(\epsilon |\cos (2\omega)|)^{-{1\over 2}}. 
\label{q20}  
\end{align}

\begin{figure*}
    \includegraphics[width=0.5\linewidth]{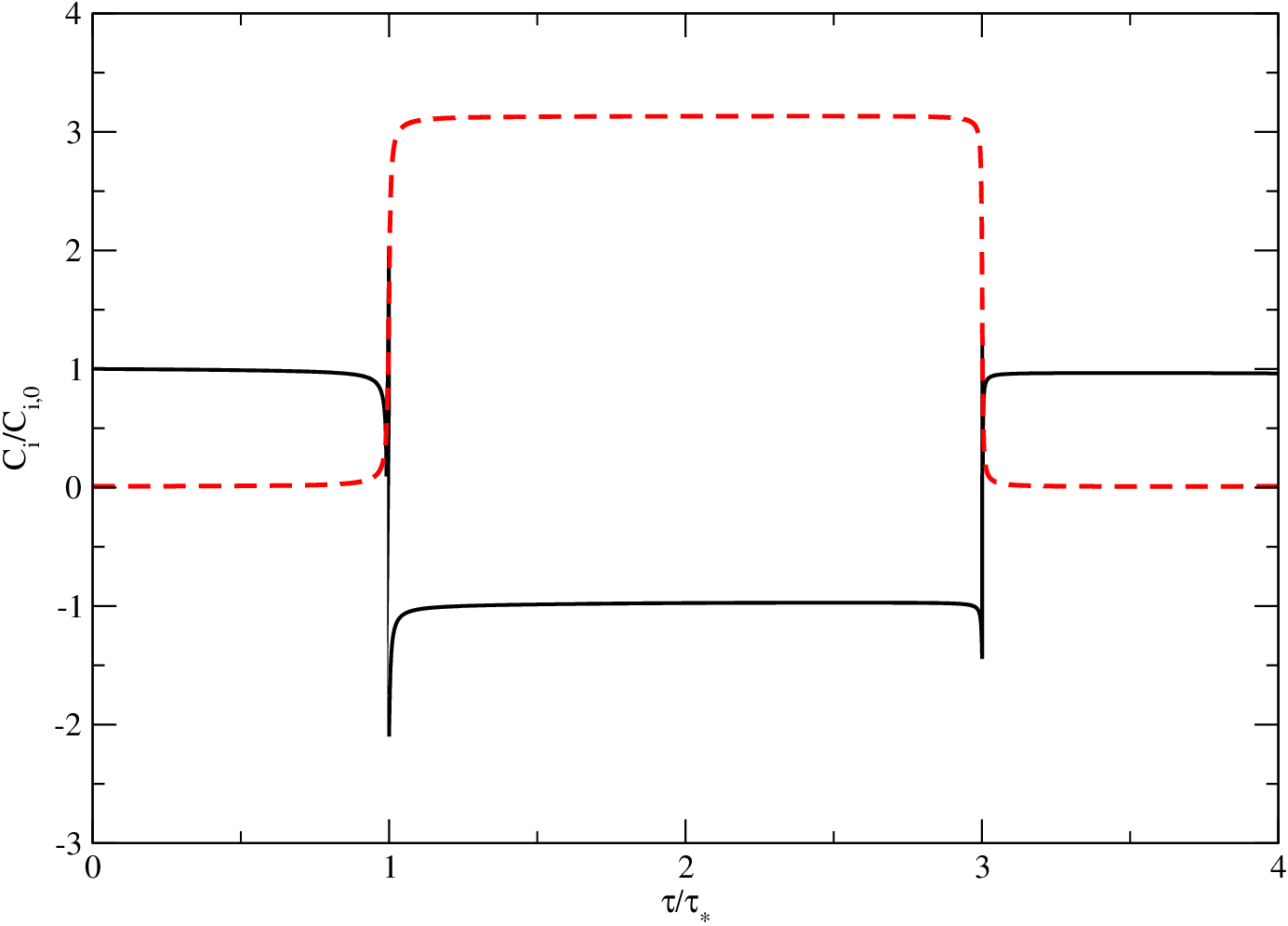}\par 
\caption{The dependencies of the ratio $C_{i}/C_{i,0}$ and $i$ on time shown as the solid line and dashed lines, respectively. 
The initial values of  $i$, $e$ and $\omega$ are the same as the ones used in Fig. \ref{Fig2}.
To plot this Figure we use $i_{+}=i$ when $i \le \pi/2$ and $i_{-}=i-\pi$ is used otherwise in eq. \ref{q20}.}
\label{Fig3}
\end{figure*}

In order to check eq. (\ref{q20}) we show the ratio $C_{i}/C_{i,0}$, where $C_i$ is evaluated for numerically calculated values 
of $i$, $e$ and $\omega$, and  $C_{i,0}$  is a value of this quantity calculated for
the initial values of these variables, $C_{i,0}\approx i_{0}\sqrt{|\cos (2\omega)|}$, in Fig. \ref{Fig3}. One can see that $C_{i}$ is indeed close to a constant when $i$ is
either close to zero or $\pi$. Moreover, when $i$ is close to $\pi$ the value of $C_{i}$ is close to the negative of its initial
value. 

\subsection{The evolution in the region of sharp jumps}

As follows from our previous analysis when $\tau \approx \tau_{*}$ $e\approx 1$ and $i$ changes abruptly from $0$ to $\pi$,
while when $\tau \approx 3\tau_*$ $e\approx 1$ again and we have the reverse abrupt change of $i$ from $\pi$ to $0$. From eqns (\ref{q12}) and (\ref{q18}) it is seen that we have $\cos \alpha \approx 0$, $\sin \alpha\approx -1$ and $\sin 2\omega \approx -1$,
$\cos 2\omega \approx 0$ in the former case and $\cos \alpha \approx 0$, $\sin \alpha\approx 1$ and $\sin 2\omega \approx 1$,
$\cos 2\omega \approx 0$ in the latter case. From these relations it is easy to see that we always have $\sin 2\varpi \approx 
0$ and $\cos 2\varpi \approx 1$ in both cases. Therefore, when studying the evolution in the transition layers we can set 
$e=1$, $\sin 2\varpi = 0$, $\cos 2\varpi =1$ and $\cos 2\omega \approx 0$, $\sin 2\omega \approx \kappa$\footnote{Here
and in eqns (41) and (42) we take into account that the change of $i$ from $0$ to $\pi$ takes place when $\kappa=-1$ and the opposite 
change corresponds to $\kappa=1$.} in r.h.s. of 
eqns (\ref{q9a}) and (\ref{q9b}) and change the time derivative of eccentricity in l.h.s of  eq. (\ref{q9a}) to the time 
derivative of $\epsilon=\sqrt {1-e^2}$ according to the rule ${d e\over d\tau}\approx -\epsilon {d \epsilon \over d\tau}$. 
In this way we obtain
\begin{align}
{d \epsilon \over d\tau}= 20\kappa \cos i, \quad  {d i \over d\tau}=-\kappa {20\sin i\over \epsilon}.
\label{q21}  
\end{align}
From these equations we easily obtain 
\begin{align}
\epsilon ={\epsilon^{\kappa}_{min} \over |\sin i|},
\label{q22}  
\end{align}
where $\epsilon^{\kappa}_{min}$.
It is clear that $\epsilon^{\kappa}_{min}$ provide the minimal values of $\epsilon$ during the evolution in the {{sharp jumps,}}
and,
therefore, their values are of interest for our purposes.

When $|\sin (i)|\ll 1$ we have from eq. (\ref{q22}) $\epsilon\approx {\epsilon^{\kappa}_{min} \over |i_{\pm}|}$. This expression should be matched to eq. (\ref{q20}), where we use  $C_{i,0}\approx i_{0}\sqrt{|\cos (2\omega)|}$ and 
should substitute the values of $|\cos (2\omega)|$ in the limit $e\rightarrow 1$ given by eq.
(\ref{q23}). 
In this way we have
\begin{align}
\epsilon^{-}_{min} =i_{0}{{(|\cos (2\omega_0)|)}^{1/2}\over 2^{1/2}{(C_{\omega}^{-})}^{1/4}}, \quad 
\epsilon^{+}_{min} =i_{0}{{(|\cos (2\omega_0)|)}^{1/2} {(C_{\omega}^{+})}^{1/4} \over 2^{1/2}}, 
\label{q24}  
\end{align}
where we assume that factors $1-e_{0}^2$ entering the quantities $C_{\omega}^{\kappa}$ given below eq. (\ref{q18}) are equal to unity. Substituting explicitly $C_{\omega}^{\kappa}$ in eq. (\ref{q24}) we finally obtain
\begin{align}
\epsilon^{\kappa}_{min} =i_{0}{{(1+\kappa \sin (2\omega_0))}^{1/2} \over 2^{1/2}e_{0}^{1/5}}. 
\label{q25}  
\end{align}
Equations (\ref{q25}) tell that, in general, the minimal values of $\epsilon$ are different in the two transition layers.
The expressions (\ref{q25}) take an especially simple form being expressed in terms
of the angle $\psi_0$
\begin{align}
\epsilon^{-}_{min} =i_{0}{|\sin(\psi_0)| \over e_{0}^{1/5}}, \quad \epsilon^{+}_{min} =i_{0}{|\cos(\psi_0)| \over e_{0}^{1/5}}.
\label{q25a}  
\end{align}

Now we substitute eq. (\ref{q22}) in eq. (\ref{q21}) and integrate the result choosing the integration constant in such a way that 
$\epsilon=\epsilon^{-}$ when $\tau=\tau_{*}$ and 
$\epsilon=\epsilon^{+}$ when $\tau=3\tau_{*}$. We have
\begin{align}
\epsilon =\epsilon_{min}^{\kappa}\sqrt{(1+{({20(\tau-\tau_{\kappa})\over \epsilon_{min}^{\kappa}})}^2)}, \quad 
\cot (i) = \kappa {20(\tau-\tau_{\kappa})\over \epsilon_{min}^{\kappa}}, 
\label{q26}  
\end{align}  
where $\tau_{-}=\tau_{*}$, $\tau_{+}=3\tau_{*}$.

\begin{figure*}
    \includegraphics[width=0.5\linewidth]{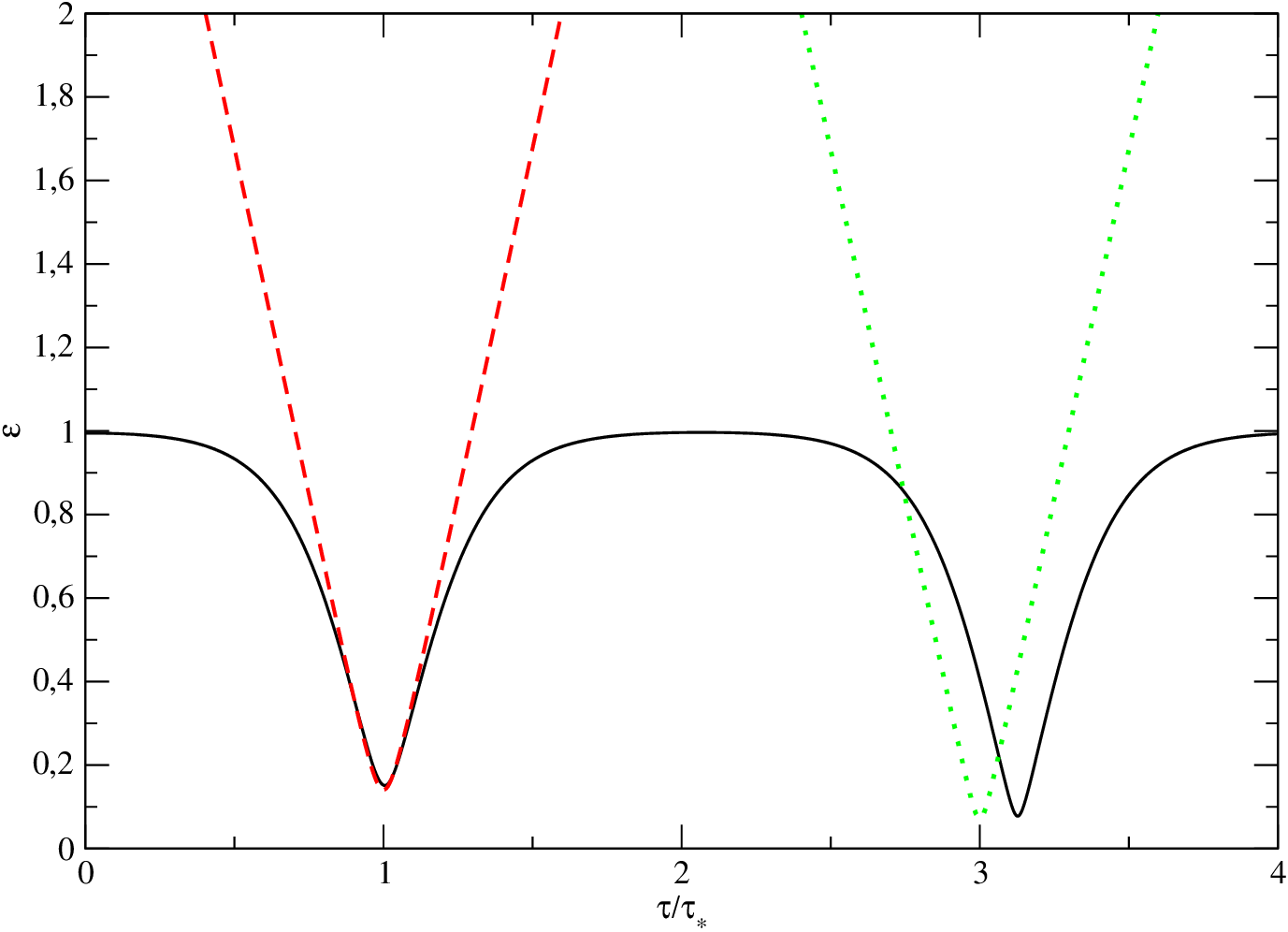}\par 
\caption{The numerically obtained dependency of $\epsilon$ on time shown as the solid line and the analytic expression given
by eq. (\ref{q26}), where the dashed line corresponds to $\kappa=-$, while the dotted line  corresponds to $\kappa=+$. 
The initial values of  $i$, $e$ and $\omega$ are the same as the ones used in Figs \ref{Fig2} and \ref{Fig3}.
}
\label{Fig4}
\end{figure*}

We show the result of comparison of the expressions (\ref{q26}) with the numerical calculation 
of $\epsilon$ in Fig. \ref{Fig4}\footnote{Note that although only one set of the initial values of our dynamical variables is represented we have checked that a similar situation takes place when other sets are used.}. One can see that the analytic expressions provide values of $\epsilon_{min}^{\kappa}$ and the shapes 
of the curves in the regions $\tau \sim \tau_{\kappa}$, which are in a quite good agreement with the numerical
results. It is interesting to see that the minimum of the numerical curve in the vicinity of $\tau =\tau_{+}$ has a rather significant shift with respect to $3\tau_{*}$. This is due to next-order corrections determined by $i_{0}$, which are not taken into account in our simple scheme. The shift gets smaller when $i_0$ is chosen to be smaller. Thus, the next order corrections lead to some evolution of the quantities,
which are assumed to be fixed, such as $\omega_0$, $i_0$, $\epsilon_{min}^{\pm}$, etc., which are significant when
a sufficiently large number of repeating cycles is considered. Taking into account the effect of Einstein precession also causes some drift of 
these quantities. This effect will be discussed in Section \ref{aps} in some detail.

\bigskip 
\section{The evolution of the dynamical system on a time scale much larger than $t_*$ 
and the role of Einstein precession}
\label{aps}

\bigskip
As we have mentioned above, the parameters describing our approximate solution slowly evolve with time when a large
number of cycles are considered. This effect is determined both by unaccounted next-order corrections to our asymptotic
solution, which is strictly valid only in the limit $e_0, i_0 \rightarrow 0$ and by additional effects causing the evolution
of the orbital elements. { Among the latter the most important role appears to be played by the Einstein apsidal precession, 
which is described by eq. (\ref{Ein1}), which should be added to the expression (\ref{q9c}).} We consider this role in this Section
assuming that the parameter $\beta_E$ in eq. (\ref{Ein1}) is small. We also briefly discuss the role played by the next-order corrections in the purely Newtonian problem, limiting ourselves to numerical means in this Paper. We also set $\pi=\pi/2$ throughout this Section, assuming, for simplicity,  that the perturbing potential is fully determined by the $|m|=2$ terms.     

Since the rate of GW emission strongly depends on the values of $\epsilon_{min}^{\kappa}$, we need to know
how the quantities entering eq. (\ref{q24}) evolve. There are some simple arguments based on the conservation of the full Hamiltonian
(\ref{Htot}), which allow us to assume that $e_{0}$ evolve slower than $\omega_0$ (or $\psi_0$) and $i_0$. Therefore,
in this Section, we consider only the evolution of latter quantities.

It is evident from eq. (\ref{Ein1}) that the Einstein precession is most important during the evolution in the region of
sharp jumps,
when $\epsilon $ is close to its maximum values during a particular cycle. This suggests the following
approximation scheme, which allows us to reduce the influence of the Einstein term to a simple map. That is, we assume that
it operates only in the transition layers, while in the regions where $i$ is close to $0$ or $\pi$ its influence
is negligible, and we can use our Newtonian asymptotic solution obtained above. However, the parameters $i_0$ and $\omega_0$ change
their values as the system evolves through sharp jumps.
The parameters are accordingly changed to new values
$i_1$ and $\omega_1$ when the system evolves from the state with $i\sim 0$ to $i\sim \pi$ ($\kappa=-1$) and after the change
of inclination from $\sim \pi$ to $\sim 0$ we have another change in the parameters $i_1, \omega_1 \rightarrow i_2, \omega_2$.
The values $i_2, \omega_2$ are then considered as initial values of these parameters for the next cycle.

\subsection{The evolution of the angles $\omega$ and $\varpi$ in the region of {{sharp jumps}} 
which takes into account the Einstein precession}

As has already been pointed out above when the system evolves in the region of sharp jumps 
$\sin (2\varpi)$ and $\cos (2\Omega)$ are
small. This property remains to be valid when $\beta_E$ is small enough. It then follows that equations (\ref{q21}) still remain approximately valid even when $\beta_E$ is non-zero and we can use their solutions (\ref{q26}) to describe the evolution of
$e$ and $i$. In order to find the evolution of $\omega$ and $\varpi$ we define $\Delta_1=\cos (2\Omega)$ and $\Delta_2=\sin (2\varpi)$, assume that the absolute values of $\Delta_{1,2}$ are small, express $\varpi$ and $\omega$ in 
eq. (\ref{q9c}) and eq. (\ref{q9d}) in terms of $\Delta_1$ and $\Delta_2$ taking into account that $2\omega\approx \kappa {\pi\over 2}-\kappa \Delta_1+2\pi k$, set $e=1$ and $\epsilon=0$ in the brackets on r.h.s. of eqns (\ref{q9c}) and (\ref{q9d})
and add eq. (\ref{Ein1}) to eq. (\ref{q9c}). In this way we obtain
\begin{align} 
{d\Delta_1\over d\tau}={20\over \epsilon}\Delta_2, \quad {d\Delta_2\over d\tau}={20\kappa \cos i\over \epsilon}\Delta_2
+{2\beta_{E}\over \epsilon^2}.
\label{q28}  
\end{align}  
We substitute eq. (\ref{q26}) in eq. (\ref{q28}) and introduce a new time variable, $T$, according to the rule 
$T={20\over \epsilon_*}(\tau-\tau_{\kappa})$,
where $\epsilon_{*}$ is any of $\epsilon_{min}^{\kappa}$. We have
\begin{align} 
{d\Delta_1\over dT}={\Delta_{2} \over \sqrt{1+T^2}}, \quad {d\Delta_2\over dT}={\Delta_2 \over (1+T^2)}+{\beta_*\over (1+T^2)},
\label{q29}  
\end{align}
where we take into account that $\epsilon=\epsilon_{*}\sqrt{1+T^2}$ in the transition layers and 
$\beta_{*}={\beta_{E}\over 10\epsilon_{*}}$. A solution to eq. (\ref{q29}) should agree with eq. (\ref{q23}), which tells us
that $|\Delta_{2}|\propto |T|$ in the limit $|T|\rightarrow \infty$. Such a solution to eq. (\ref{q29}) has the form
\begin{align} 
\Delta_1= (C+\beta_{*})T+\beta_{*}\sqrt{(1+T^2)}, \quad \Delta_2= (C+\beta_{*})\sqrt{(1+T^2)}+\beta_{*}T. 
\label{q30}  
\end{align}
The constant $C$ should be matched to the expressions (\ref{q18b}-\ref{q23}) taken in the limit $e\rightarrow 1$ and $\epsilon \rightarrow 0$, where we should use $i_0$ and $\omega_{0}$ when $\kappa=-1$ and $i_{1}$ and $\omega_{1}$  when $\kappa=1$, as explained above. It is evident from eq. (\ref{q30}) that
\begin{align} 
\Delta_1 (T\rightarrow -\infty)= CT, \quad \Delta_1(T\rightarrow \infty)=(C+2\beta_{*})T. 
\label{q31}  
\end{align}   
Thus, leaving aside the short time evolution of the state variables in the region of sharp jumps,
one can see that 
the effect of Einstein precession leads to redefinition of the constant $C$, which, in its turn, leads to 
redefinitions of the angles $\omega_0$ and $i_0$ as has been  pointed out above. That allows us to describe the influence of Einstein precession in terms of an iterative map linking 'old' and 'new' values of  $\omega_0$ and $i_0$. This map is analyzed in Appendix \ref{map}. As shown there this map has the property that one of $\epsilon^{\kappa}$
always grows, while another one always decreases with the number
of its iterations $n$.

\subsection{An estimate of a minimal value of $\epsilon$ in case when $\beta_{E}$ is relatively large}

From equations (\ref{q42}-\ref{q46}) of Appendix \ref{map}
it follows that $i_0$ as well as either $\epsilon^{-}$
or $\epsilon^{+}$ tend to infinity with $n$. This clearly violates our basic assumptions, and,
therefore, our treatment is valid only for a finite value of cycles. As a criterion of violation
of our assumption we suppose that they are broken when $\Delta_1=\cos (2\omega)$ is order of one
when $i=\pi/2$. Remembering that the latter condition corresponds to $T=0$ we have from equation (\ref{q30})
$\Delta_1(i=\pi/2)=\beta_{*}={\beta_{E}\over 10\epsilon_*}$, and, from the condition $\Delta_1 < \beta_{*}$ 
we obtain
\begin{align} 
\epsilon_{*} > \epsilon_{min}={\beta_E\over 10},
\label{q49}  
\end{align}
and, using either eqns (\ref{q45}) and (\ref{q46}) we conclude that our map is expected to be valid 
when $n < n_{max}$, where 
$n_{max} \approx \Lambda ({5{i_{0}(0)}^2|\sin (2\psi_0(0))|\over \beta_E})$ 
and
$\Lambda = \log ({10 i_{0}(0)|\sin \psi_{0} \over e_{0}^{1/5}\beta_{E}})$ 
when 
$\sin (2\psi_0(0)) > 0$
and 
$\Lambda = \log ({10 i_{0}(0)|\cos  \psi_{0} \over e_{0}^{1/5}\beta_{E}})$ 
in the opposite case.

The inequality (\ref{q49}) also follows from another quite general argument. Namely, 
we can estimate a minimal 
possible value of $\epsilon$, which can in principle be reached when $\beta_{E}$ is non-zero using the
fact that Hamiltonian is the integral of motion. For that, we note 
that from eq. (\ref{Ein2}) it follows that the contribution to the Hamiltonian due to the presence of Einstein
precession is always negative, and, on the other hand, it follows from eq. (\ref{q7}) that the corresponding terms 
due to the tidal terms can be positive. It is possible to show that when $e_{0}$ and $i_{0}$ are small,
but, on the other hand we consider the state of the system with $e\sim 1$ and $i \sim \pi/2$, their 
separate absolute values are much larger than a value of Hamiltonian, so they should compensate each other. Setting $e=1$ and
 $i = \pi/2$  in eq. (\ref{q7}), and assuming that the value of the total Hamiltonian is equal to  zero we obtain
\begin{align} 
5\cos (2\omega )(\cos (2\varpi )+1)-{\beta_{E} \over \epsilon}\approx 0.
\label{q50}  
\end{align}
The first term in eq. (\ref{q50}) cannot be larger than $10$, which corresponds to $\cos (2\omega)=\cos (2\varpi)=1$.
Setting it equal to $10$ we obtain the expression for $\epsilon_{min}$.

\begin{figure*}
    \includegraphics[width=0.5\linewidth]{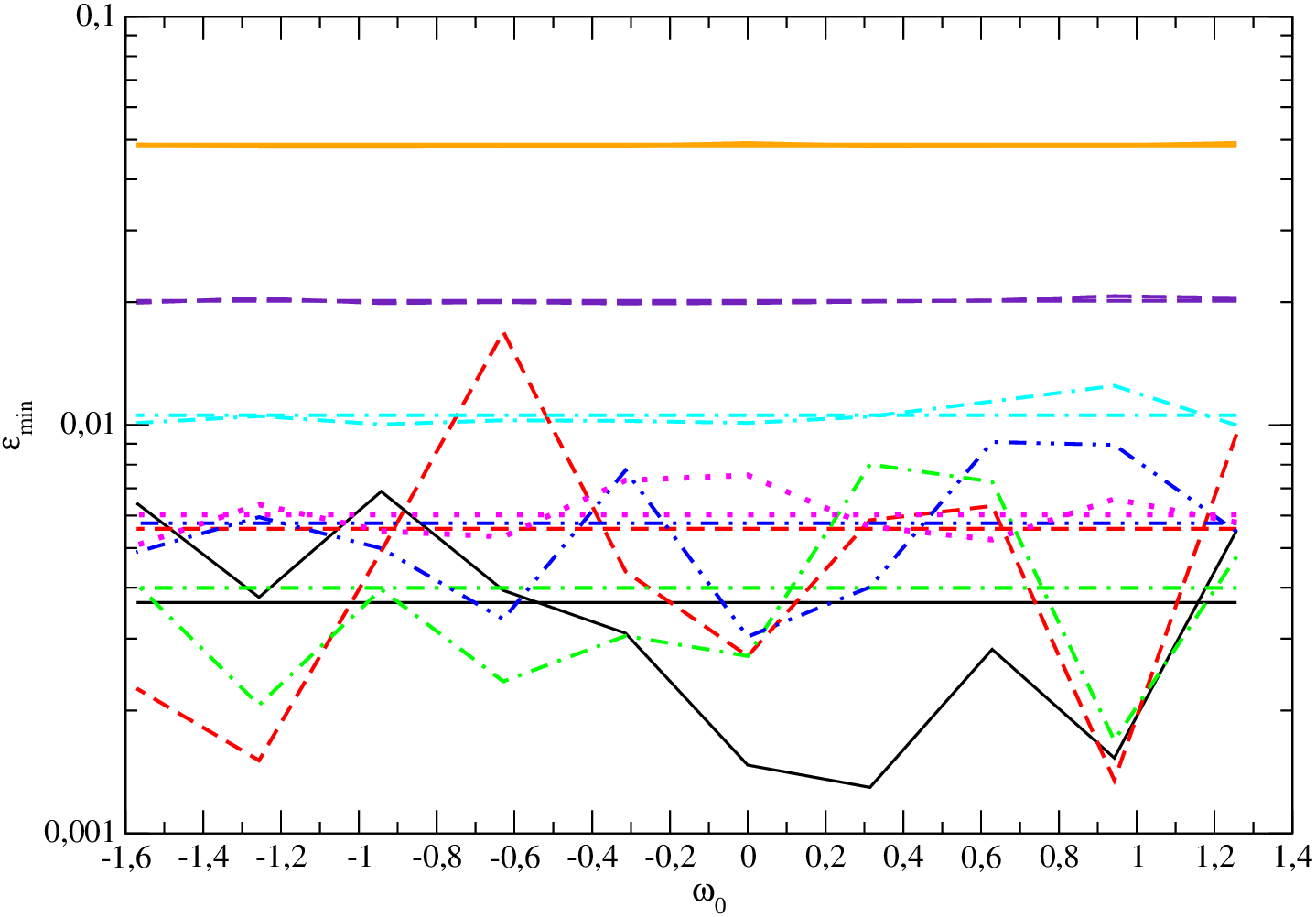}\par 
    \includegraphics[width=0.5\linewidth]{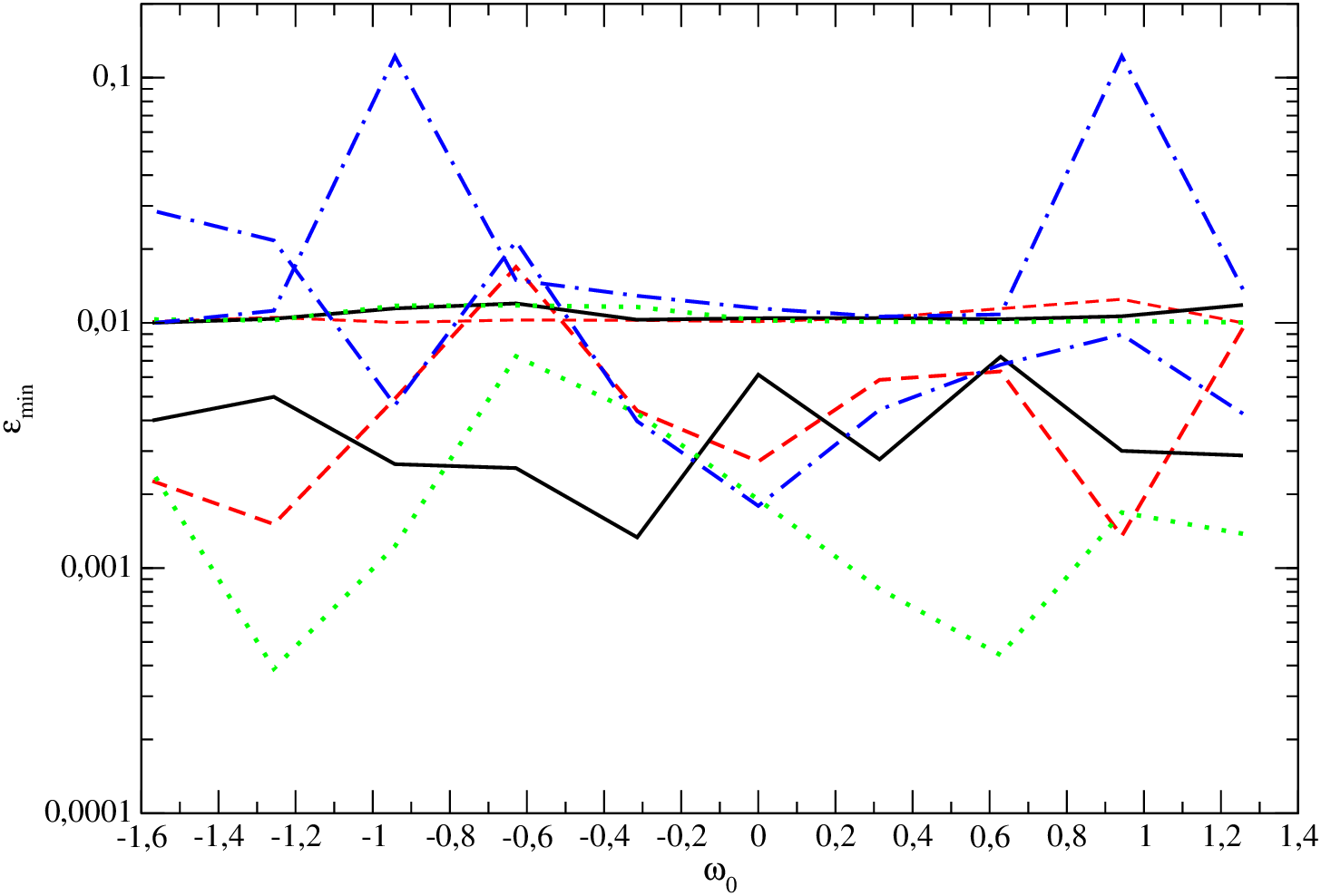}\par 
\caption{Top panel. The minimal values of $\epsilon_{min}$ found numerically over the 
calculation time corresponding to $10^2$ cycles ($\sim 400\tau_{*}$) as functions of the initial
nodal angle $\omega_{0}$. $e_{0}=i_{0}=0.1$ for all runs. Different curves correspond to different values 
of $\beta_E$ with the horizontal lines of the same style showing the averaged over $\omega_0$ values of
$\epsilon_{min}$. The solid curves and horizontal lines represent { $\beta_{E}=0$ (black)} and the largest considered { $\beta_{E}=0.5$ (orange)},
with the larger values corresponding to { $\beta_E=0.5$}. Note that in the latter case $\epsilon_{min}$ practically coincides with its average value.  The short dashed, dot dashed, double dot dashed, dotted, dot double dashed, long dashed
curves are for { $\beta_E=10^{-3}$ (red), $0.1$ (green), $0.025$ (blue), $0.05$ (magenta), $0.1$ (cyan) and $0.2$ (violet)}, respectively. Bottom panel. The same as
top panel, but different curves correspond to different values of $i_0$. $e_0=0.1$ for all curves. The curves
of the same style with larger (smaller) values  
correspond to { $\beta_E=0.1$ ($\beta_{E}=0.001$)}. { Solid
black, dashed red, dotted green and dot dashed blue curves} are for $i_{0}=0.05$, $0.1$,
$0.2$ and $0.5$, respectively.}
\label{Fig8}
\end{figure*}

\begin{figure*}
    \includegraphics[width=0.5\linewidth]{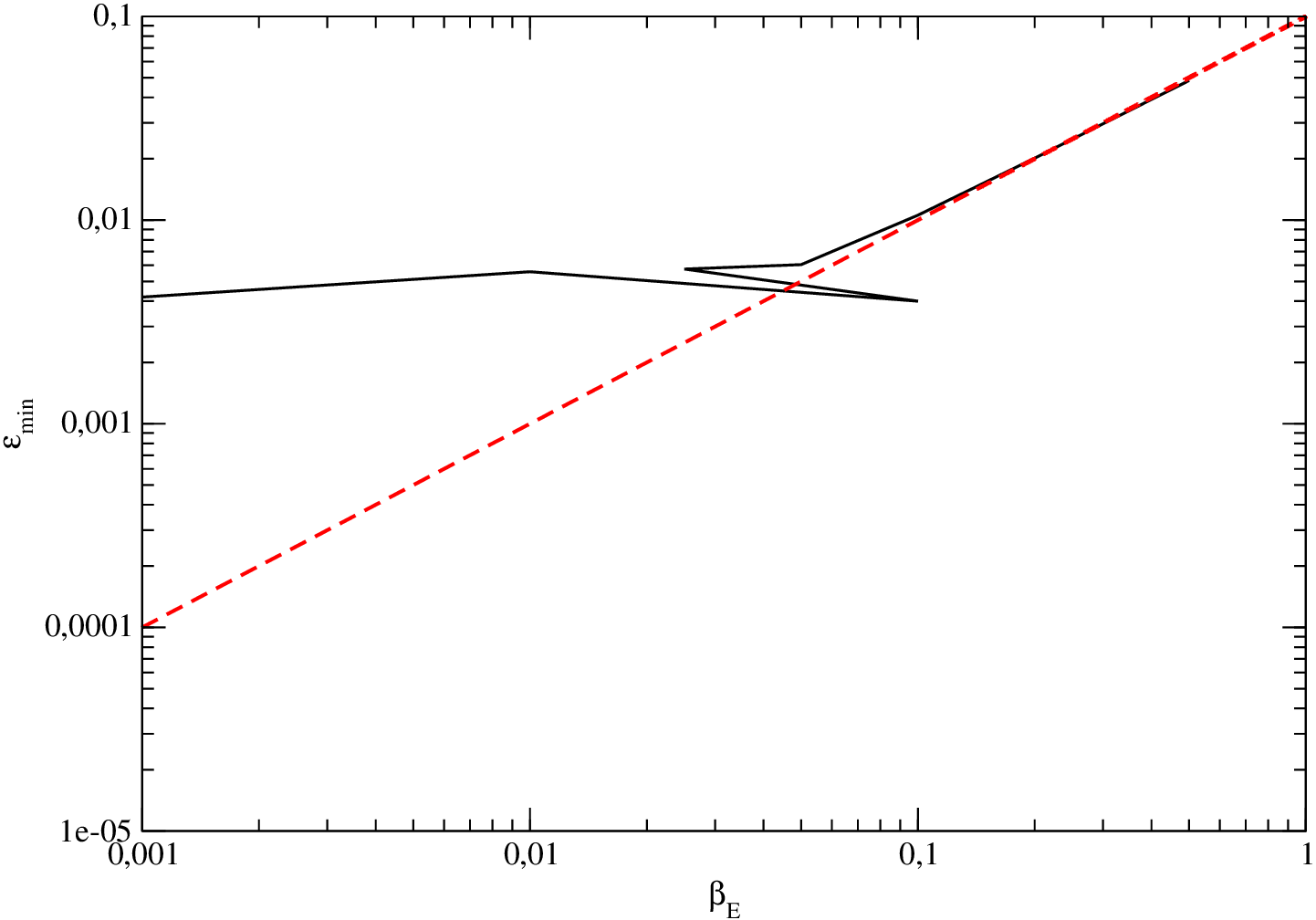}\par 
    \includegraphics[width=0.5\linewidth]{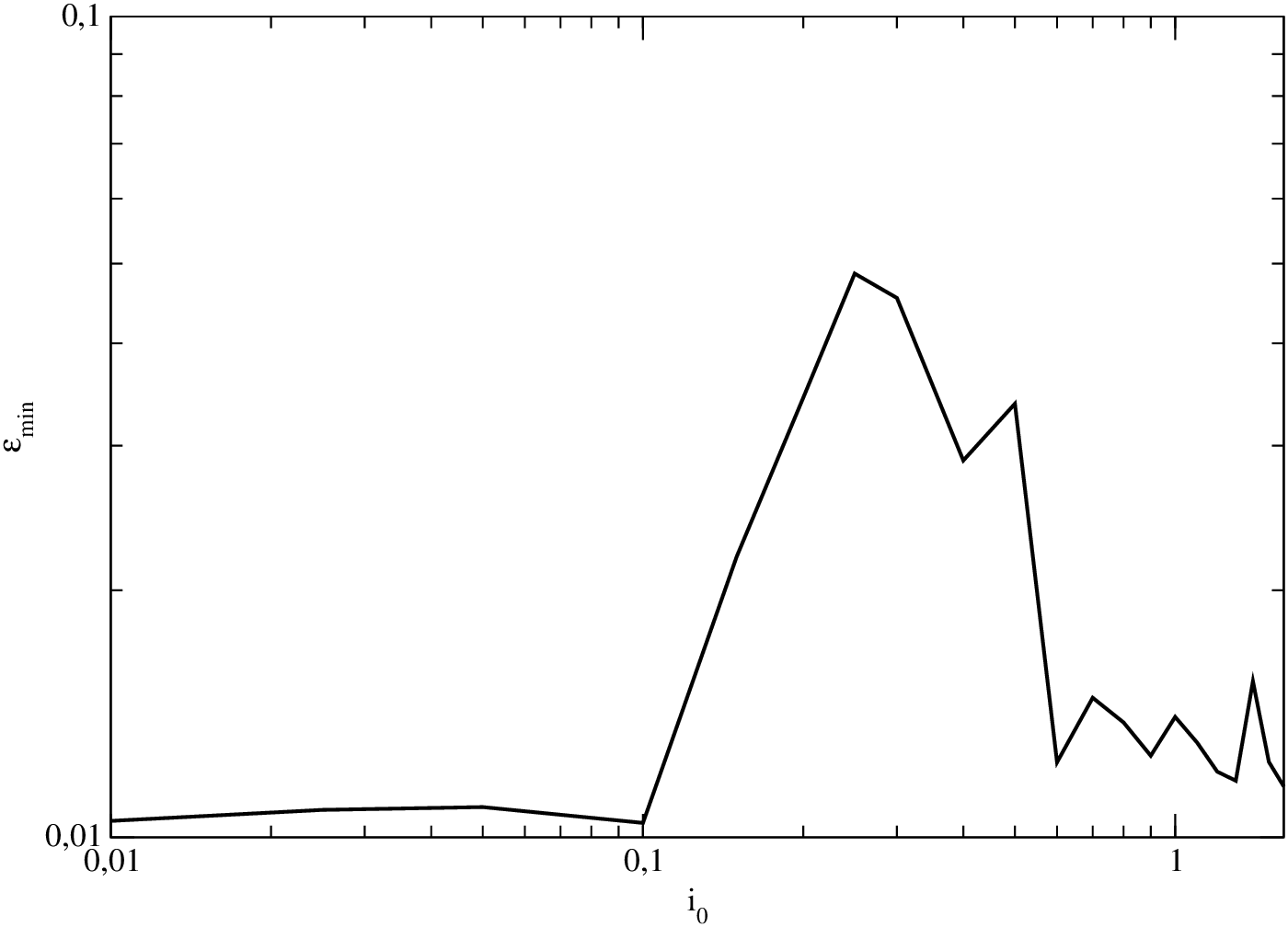}\par 
\caption{Top panel. The value of $\epsilon_{min}$ averaged over the initial nodal angle $\omega_0$ as a function of $\beta_{E}$ is
shown as a solid line. $e_0=i_0=0.1$ for the shown case. The dashed line represents the theoretical value (\ref{q49}). Bottom panel. The same quantity
calculated numerically as a function of $i_0$. We use $e_0=\beta_E=0.1$ 
for all calculations.}
\label{Fig9}
\end{figure*}

\begin{figure*}
    \includegraphics[width=0.5\linewidth]{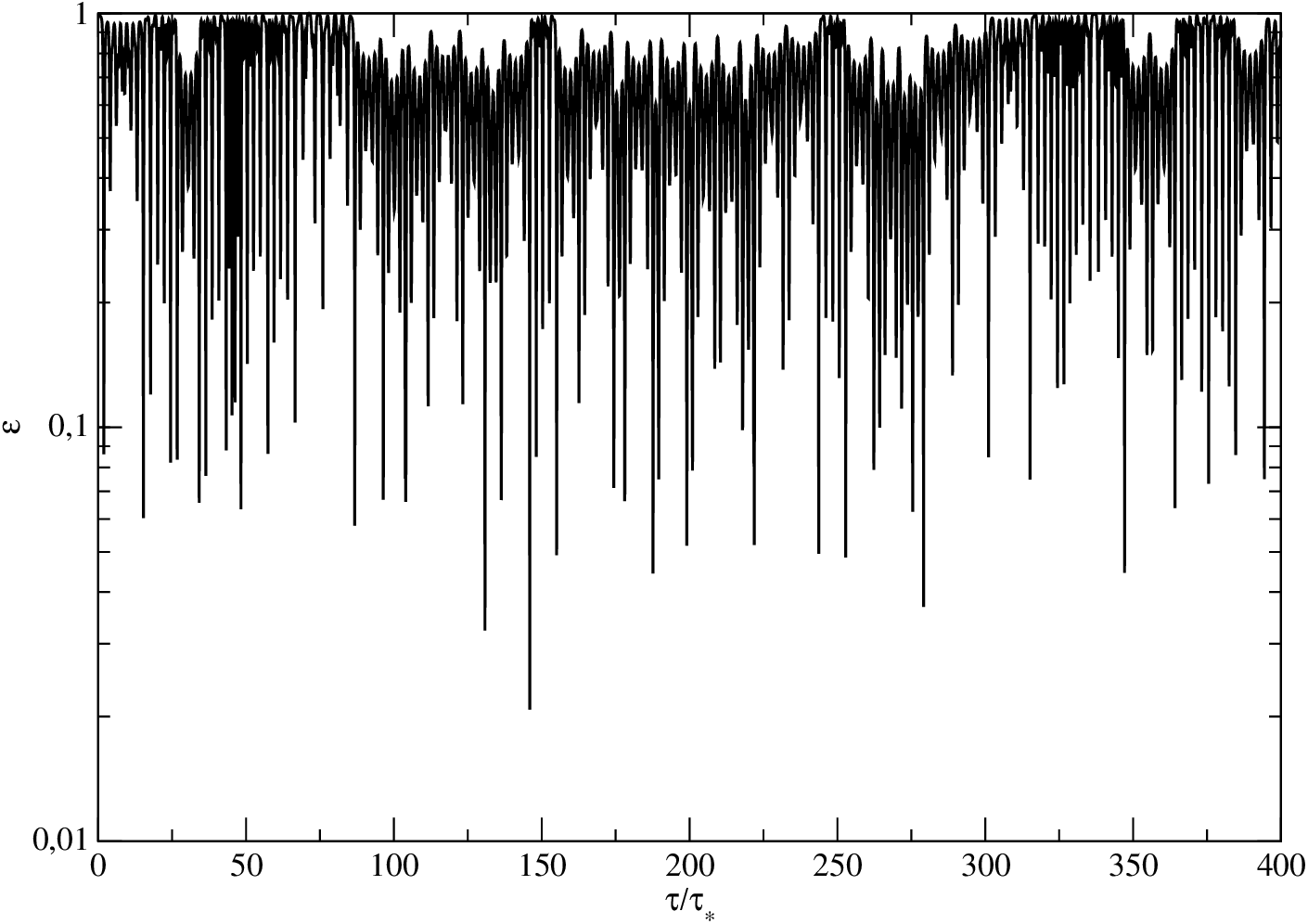}\par 
    \includegraphics[width=0.5\linewidth]{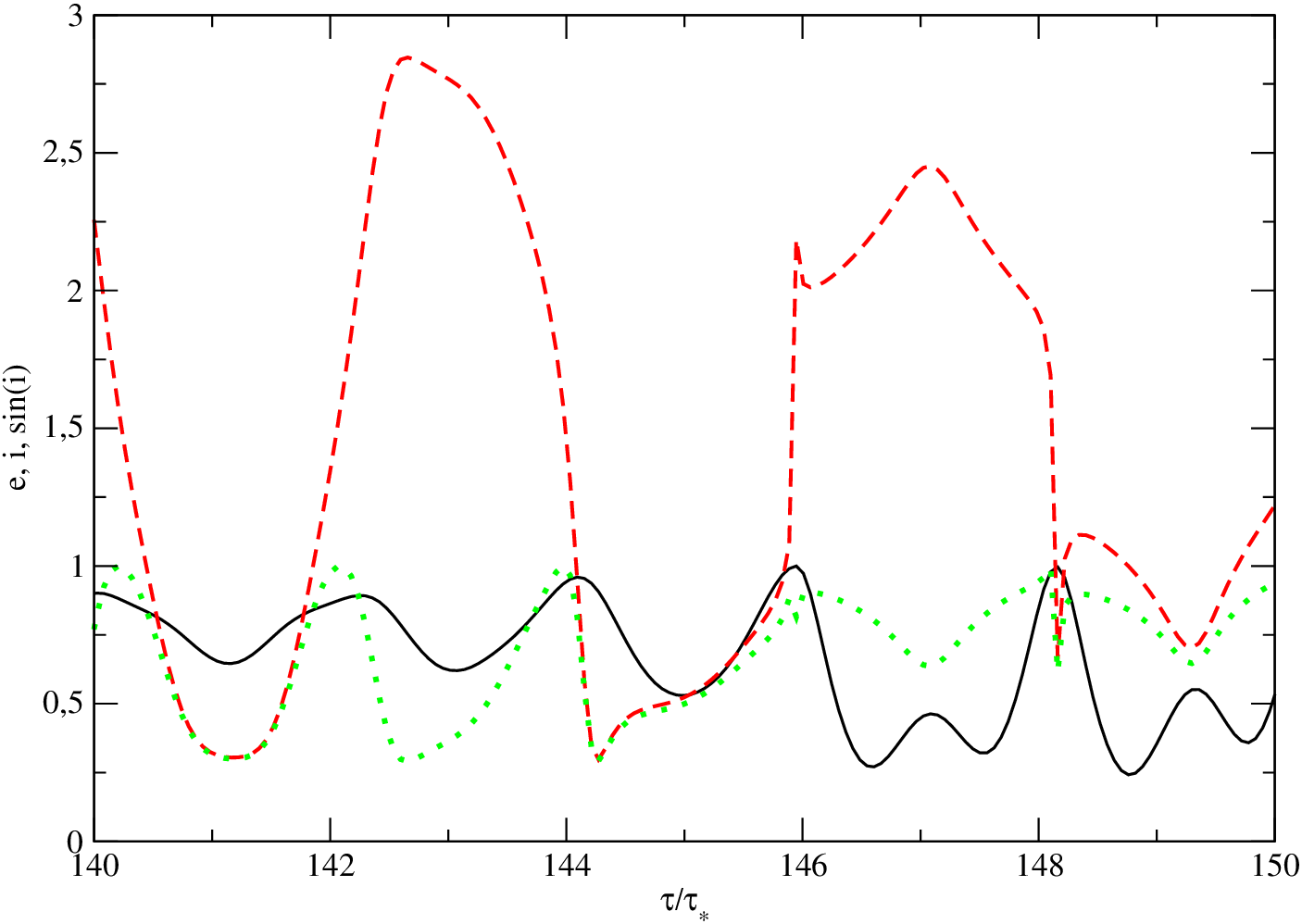}\par 
\caption{Top panel. The value of $\epsilon_{min}$  as a function of time
numerically obtained for the calculation with $e_0=\beta_E=0.1$, $i_0=\pi/4$ and $\omega_0=\pi/3$. We show the whole time period $t < t_{end}$. 
Bottom panel. $e$ (solid line), $i$ (dashed line) and $\sin (i)$ (dotted line) as functions of time shown in the time interval $140 < \tau/\tau_{*} < 150$.}
\label{Fig10}
\end{figure*}   

\subsubsection{A minimal value of $\epsilon_{*}$ obtained from a numerical analysis of our system} 

In order to check the analytic approach discussed above we perform a set of numerical runs with different initial 
parameters. The initial eccentricity $e_0$ is set to be equal to $0.1$ for all runs, but we consider several initial inclinations $i_0$ in the range $0.05-1.6$, the parameter $\beta_E$ in the range $0-0.5$ and ten values of the initial
nodal angle $\omega_0$ uniformly distributed in the interval $(-\pi/2..\pi/2)$ \footnote{We remind that we always set 
initial value of the apsidal angle to be equal to the negative of the nodal angle.}. The end time of the computation, 
$t_{end}$, is chosen in such a way that approximately 
$100$ evolutionary cycles are executed in a particular computational run. From the results discussed above it follows that every cycle has a period $\approx 4\tau_*$, where $\tau_{*}$ is
given by eq. (\ref{q15}), and, therefore 
We end up the computations when  $t=t_{end}=400\tau_{*}$. From equations (\ref{q8n}) and (\ref{q15}) we see that $t_{end}$ in the physical units can 
be represented as
\begin{align} 
t_{end}\approx 160\log ({8\over e_0^2}){n_0\over \Omega^2}\approx 10^{3}{n_0\over \Omega^2},
\label{q51}  
\end{align} 
where we set $e_0=0.1$ to obtain the last equality. We obtain minimal values of $\epsilon$ for each run and 
compare them with our analytic estimates.

The results of the comparison are shown in Figs. \ref{Fig8} and \ref{Fig9}. In top panel of Fig. \ref{Fig8} we plot the
so obtained numerical values of $\epsilon_{min}$ versus $\omega_{0}$ for our 'standard choice' of the initial eccentricity
and inclination $e_0=i_0=0.1$. Curves of different style and color correspond to different values of $\beta_E$ and we
also show the averaged over $\omega_0$ values of $\epsilon_{min}$ as horizontal lines of the same style and color, see
the figure caption for a description of particular lines. As seen from this plot when $\beta_E > \sim 0.05$ the values 
of $\epsilon_{min}$ quite weakly depend on $\omega_{0}$. At smaller values of $\beta_{E}$ the curves are less uniform, but
particular values of $\epsilon_{min}$ deviate from the averaged one by factor of a few even in this case. In bottom panel
of the same figure we plot $\epsilon_{min}$ corresponding to different initial inclinations as functions of $\omega_0$ for $e_0=0.1$ and two representative values of $\beta_{E}=10^{-3}$ and $0.1$. As seen from this plot, in general, the curves
with different values of $i_0$ do not deviate much from each other when $i_0$
is sufficiently small.

In Fig. \ref{Fig9} we show the averaged value of $\epsilon_{min}$ as a function of $\beta_{E}$ for $e_0=i_0=0.1$ in top panel and the same quantity as
function of $i_0$ for $e_0=\beta_{E}=0.1$ in bottom panel. In top panel
the solid
curve represents the numerical results, while the dashed line shows our expression for $\epsilon_{min}$ given in eq. (\ref{q49}). As seen from this Fig. when $\beta_{E} > 0.05$ we have an excellent agreement between the numerical results
and our analytic expression. In this opposite case $\beta_{E} < 0.05$  $\epsilon_{min}$ is almost constant, while 
when $\beta_{E} \sim 0.05$ the numerical curve shows a complicated behaviour. In general, the numerical curve agrees
with our previous analysis. When $\beta_{E}$ is significantly large the effect of Einstein precession gives the main
contribution to the evolution of the approximate constants of motion characterizing our approximate asymptotic solution
discussed in the previous Section, and, accordingly, to the evolution of $\epsilon_{min}$ from one cycle to another. In
the opposite case this role is played by non-linear corrections to the asymptotic solution, which do not depend on
$\beta_{E}$. We have, accordingly, an almost constant dependency of $\epsilon_{min}$ on $\beta_{E}$ in the latter case.
In what follows we are going to use a very simple approximation of this dependency. Namely, we are going to assume that
\begin{align} 
\epsilon_{min}(\beta_{E} > 0.05)={\beta_{E} \over 10}, \quad \epsilon_{min}(\beta_{E} < 0.05)=5\cdot 10^{-3}. 
\label{q52}  
\end{align}   

Bottom panel of this Fig. shows that the dependency of the averaged {{$\epsilon_{min}$ on $i_0$}} is non-monotonic. It agrees with our theoretical value
$\epsilon_{min}(\beta_{E}=0.1)=0.01$ only when $i_0 < \sim 0.1$. When $i_0$ gets larger $\epsilon_{min}$ increases, then it decreases back to values $\sim 0.011-0.15$. 

Interestingly, the average value of $\epsilon_{min}$ is quite small even when $i_{0}\approx \pi/2$. This is important for our purposes, since it greatly extend an available phase 
space of orbits with small eccentricities, which can experience an efficient circularization due to GW emission. The case of large initial inclinations cannot be treated, however, in framework of the simple theory developed above.  

In order to illustrate the behaviour of our system in case of a large $i_{0}$ we
show time the time dependency of $\epsilon$ for the whole time interval $0 < t < t_{end}$
in Fig. \ref{Fig10}, top panel for $i_0=\pi/4$ and $\beta_E=0.1$. As seen from this Fig. the values of $\epsilon$ can be as small as $\sim 2\cdot 10^{-2}$ in this particular case. A smaller
time interval containing the moment of time $\tau \approx 144\tau_{*}$ 
corresponding to the smallest value of
$\epsilon$ is shown in bottom panel of the same Fig. We show the dependencies of $e$,
$i$ and $\sin (i)$. The dynamics of the variables at the time $< \sim 146$ is in qualitative agreement with our previous results. The minimal values of $i$ are about $0.3$
and the minimal values of eccentricity are about $0.5-0.6$, the maximal values of eccentricity are reached at times, which are close to the times corresponding to
$\sin (i)=1$ ($i=\pi/2$) and the cycle period is close to $4\tau_{*}$. Therefore, it
is possible to have a transient dynamics with relatively small minimal values of 
inclinations and intermediate values of eccentricity even when the initial value of
inclination is large.

\bigskip
{\section{Final stages of SBBH evolution and emission of GWs} }

If at final stages of evolution two components in SBBH are close enough emission of GWs could play important role in the evolution of SBBH at these stages. In this Section we give an order of magnitude estimations of SBBH parameters when such situation  
takes place.
\label{est}

\bigskip

For this purpose, it is enough to leave  
in the perturbing potential (see equation (1)) the terms with  
$|m|=2$ only. 

In what follows we use {{
empirical relationships}} between the effective radius of NSC, the mass of the central primary black hole, {{$M_{p}$}}, and the radius of influence of the primary, $r_{infl}$, proposed in \cite{Georg}. For simplicity, we consider only the early-type galaxies; {{also see \cite{Georg}. As follows 
from Fig. 8 of this Paper
the effective radius of such  
galaxies is of the same order as  $r_{infl}$. For this reason and for reason of simplicity, 
we assume that
they these radii are { close} to each other.  
Then after simple manipulations with some parameters presented in Table 1 of this Paper we can see that}}  
\begin{align} 
r_{infl}\approx 23 \alpha_r M_{8}^{0.35}pc,
\label{q53}  
\end{align} 
where $M_{8}=M_{p}/(10^{8}M_{\odot})$ and { the parameter $\alpha_r$
is expected to be of the order of unity}. Then, assuming that inside $r_{infl}$ the stellar density, $\rho_{st}(r)$, follows the
well known 
profile $\rho_{st}(r)\propto r^{-3/2}$ 
\cite{Young}, we have
\begin{align} 
M_{st}(r < r_{infl})\approx M_{p}({r\over r_{infl}})^{3/2},
\label{q54}  
\end{align}
where $M_{st}(r)$ is stellar mass in NSC within a given radius $r$.   
If (as we have already mentioned in the Introduction) the mass ratio of the black holes, $q=M_{s}/M_{p}$,
is small,  
dynamical friction is { expected to be} efficient 
until the SBBH semi-major axis, $a$, is larger than a scale where the stellar mass enclosed within the
orbit is { of the order of} the mass of the secondary black hole, $M_{s}$. { Introducing a new parameter $\alpha_s$ such as
$M_{st}(r=a)=\alpha_s M_s$, 
from eq. (\ref{q54}) we can see that  
the latter equation is satisfied} when 
\begin{align} 
a \approx 4.6\cdot 10^{-2}\alpha_s^{2/3}q_{-2}^{2/3}r_{infl}=1.1\alpha_{r}\alpha_{s}^{2/3}q_{-2}^{2/3}M_{8}^{0.35}pc,
\label{q55}  
\end{align}
where $q_{-2}=10^{2}q$. 

We assume that when the semi-major axis is approximately equal to the expression
(\ref{q55}) the stars inside the orbit are quickly dispersed by the secondary and there is no significant inflow of stars to the radii $r\sim a$ from the outer part of the system. {{Consequently,
the evolution of $a$ is supposed to stop when its value is approximately equal
to (\ref{q55}).}} The orbital motion is then mainly
determined by the Newtonian gravitational field of the primary at relatively short timescales,
and at larger timescales the secular evolution
of the orbital elements is assumed to be determined by the tidal field of non-spherical NSC
and the Einstein precession of the apsidal line as discussed above as well as the effect of GW emission
in the case of highly eccentric orbits.  

Typical values of the  mean motion $n_0=\sqrt{GM_p\over a^{3}}$ and the orbital period $P_{orb}=2\pi/n_0$ can be directly estimated
from eq. (\ref{q55}) as
\begin{align} 
n_0=1.8\cdot 10^{-3}\alpha_r^{-3/2}\alpha_s^{-1}q_{-2}^{-1}M_{8}^{-0.025}yr^{-1}, \quad P_{orb}=3.4\cdot 10^{3} \alpha_r^{3/2}\alpha_s q_{-2}M_{8}^{0.025}yr.
\label{q56}  
\end{align} 

In order to estimate typical values of the timescales $t_{*}$ and $t_{end}$ characterizing the secular evolution due to the tidal effects and given by eqns (\ref{q8n}) and (\ref{q51}), respectively,
we need to parametrize the frequency $\Omega$ determining the strength of the tidal potential
(\ref{q1}) in a convenient way. We represent it in the form
\begin{align} 
\Omega=\sqrt{\mu GM_p\over r_{infl}^3},
\label{q57}  
\end{align} 
where the dimensionless coefficient $\mu$ determines a degree of non-sphericity of NSC. It is
small when NSC is only slightly non-spherical. We substitute eq. (\ref{q53}) in eq. (\ref{q57}) and 
eq. (\ref{q57}) in eqns  (\ref{q8n}) and (\ref{q51}) and, taking into account eq. (\ref{q56}) we obtain
\begin{align} 
t_{*}\approx 1.9\cdot 10^{7}\alpha_r^{3/2}\alpha_{s}\mu^{-1}q_{-2}^{-1}M_{8}^{0.025}yr, \quad t_{end}\approx 250t_{*} = 4.9\cdot 10^{9}\alpha_r^{3/2}\alpha_{s}\mu^{-1}q_{-2}^{-1}M_{8}^{0.025}yr.
\label{q58}  
\end{align} 

{{It is convenient to represent 
relativistic parameter 
$\beta_{E}$, defined in eq. (\ref{Ein1}), in terms of dimensionless variable $u=GMc^{-2}a^{-1}$ which is proportional to  
the ratio of the gravitational 
radius of primary black hole
to the 
semi-major axis of secondary black hole orbit. Then taking to account eq. (\ref{q55}) we obtain
\begin{align} 
u={GM\over c^2 a}= 4.4\cdot 10^{-6}\alpha_r^{-1}\alpha_s^{-3/2}q_{-2}^{-2/3}M_{8}^{0.65}.
\label{q59}  
\end{align}
Using this definition we represent the parameter $\beta_{E}$  as $\beta_{E}= 12 u \Omega^2 n_0^{-2}$,
and, using eqns (\ref{q55}), (\ref{q57}) and (\ref{q59}) we obtain
\begin{align} 
\beta_E=0.54\alpha_r^{-1}\alpha_s^{4/3}\mu^{-1}q_{-2}^{-8/3}M_{8}^{0.65}.
\label{q60}  
\end{align}
According to eq. (\ref{q60}) 
for the considered typical parameters of the binary and NSC the effect associated with the Einstein
precession is relatively large, and, therefore, this effect should be definitely taken into account.}}  

{{Let us consider now the  
possibility of the orbital circularization due to emission of gravitational waves by SBBH 
with $\epsilon \ll 1$.  We can use the result of \cite{Peters} for the evolution of semi-major axis  
for orbits with
a small mass ratio $q$ and eccentricity close to one:
\begin{align} 
({\dot a \over a})_{GW}\approx -{170\over 3} {q u^{5/2}n_0\over \epsilon^7}.
\label{q61}  
\end{align}}}
As a criterion of an efficient orbital circularization we use the condition that the associated characteristic evolution time
$t_{GW}= |({a \over \dot a})_{GW}|^{-1}$ is smaller than the orbital period $P_{orb}$. This results in the condition {{
 \begin{align} 
q_{-2} > {3\over 3.4 \pi} \alpha_r^{5/2}\alpha_s^{5/3}\epsilon^{7}u^{-5/2}.
\label{q62}  
\end{align}
}}
Both dimensionless factors {{$u^{5/2}$}} and $\epsilon^7$ are quite small. In order to single out the small numerical factor
in $\epsilon^7$ we normalize it to the smallest expected value $5\cdot 10^{-3}$ (see eq. (\ref{q52}) using {{$\tilde \epsilon =\epsilon/(5\cdot 10^{-3})$}}. Substituting $\epsilon^7 = 7.8\cdot 10^{-17}{\tilde \epsilon}^7$ and eq. (\ref{q59}) in eq.
(\ref{q62}) we obtain
\begin{align} 
\tilde \epsilon < \epsilon_{max}=2.9 \alpha_r^{-5/14}\alpha_s^{-5/21}q_{-2}^{-0.095} M_{8}^{0.23}.
\label{q63}  
\end{align} 
The inequality (\ref{q63}) can be satisfied when $\tilde \epsilon > 1$, hence, inequalities (\ref{q63}) and (\ref{q49})  are compatible, and, accordingly, we can use $\epsilon \approx \beta_{E}/10$ { , and obtain from eq. (\ref{q60})}
\begin{equation}
\tilde \epsilon \approx 54 \alpha_r^{-1}\alpha_s^{4/3}(0.1/\mu ) q_{-2}^{-8/3}M_{8}^{0.65}.
\label{q63a}
\end{equation}
Using this estimate we can reformulate eq. (\ref{q63}) as 
\begin{align} 
q_{-2} > ~ 3.1 \alpha_{r}^{-0.25}\alpha_s^{0.61}({0.1\over \mu})^{0.38}M_{8}^{0.16}.
\label{q64}  
\end{align} 
Note a rather weak dependence on the quantities on r.h.s.

On the other hand, eq.(\ref{q52}) tells that we assume { , for our estimates,} that $\tilde \epsilon $ cannot be smaller than one. Setting $\tilde \epsilon=1$ in l.h.s of 
eq. (\ref{q63a}) we have
\begin{align} 
M_{8}  >  \sim 2\cdot 10^{-3}\alpha_r^{1.5}\alpha_s^{-2}{({\mu \over 0.1})}^{1.5} q_{-2}^{4}.      \label{q65}  
\end{align} 
Thus, for a given mass ratio { our approach is valid} only for 
sufficiently massive  
primary black holes. 

Using the Newtonian relationship between the orbital periastron $r_p$ and the semi-major axis,
$r_{p}=(1-e)a$, we can easily estimate  from eq. (\ref{q59}) a typical value of periastron 
corresponding to SBBHs, which evolve 
due to 
the emission of GWs.
\begin{align} 
r_p\approx 11.2 \alpha_r \alpha_s^{2/3} q_{-2}^{-2/3}M_{8}^{-0.65}{\tilde \epsilon}^2 {GM_p\over c^2}=4.1 \alpha_r^{5/4} \alpha_s^{1/3}({\mu \over 0.1})^{1/4}M_{8}^{-0.81}{\tilde \epsilon}^2 {GM_p\over c^2},                   
\label{q66}  
\end{align} 
{ where we use eq. (\ref{q63a}) to express the mass ratio in terms 
of other parameters of the problem.}
{ Thus, when all quantities in the last equality on r.h.s. have their nominal values,
$r_p$ turns out to be close
the radius of 
the marginally bound orbit
for a Schwarzschild primary black hole $r_{mb}={4GM\over c^2}$, which is the minimal possible periastron for parabolic orbits. Those orbits, which have 
periastra formally smaller than $r_{mb}$ directly plunge into black hole. Therefore,
our model implies that, for certain system's parameters, there could be processes of direct black holes collisions from parabolic orbits, or, in
case when $r_p$ is larger than but comparable to $r_{mb}$, there could be
a substantial eccentricity during the whole merger process.}

{ In order to illustrate this property of our model  
let us formally assume that effects of General Relativity are small, and the  relation between eccentricity and semi-major axis during the orbital evolution caused by GW emission is given by equation (5.11) of \cite{Peters} valid in the limit of weak gravity,
\begin{align} 
a(e)=c_0f(e), \quad f(e)={e^{12/19}\over (1-e^2)}(1+{121\over 304}e^2)^{870/2299},      
\label{q67}  
\end{align} 
To express $c_0$ in terms of $r_p$ given by 
eq. (\ref{q66})
 we consider the limit $e \rightarrow 1$ in eq.
(\ref{q67}) and obtain that 
\begin{align}
 c_{0}\approx {2r_p\over (1+{121\over 304})^{870/2299}}\approx 19.7\alpha_r \alpha_s^{2/3} q_{-2}^{-2/3}M_{8}^{-0.65}{\tilde \epsilon}^2{GM_p\over c^2}=7.2 \alpha_r^{5/4} \alpha_s^{1/3}({\mu \over 0.1})^{1/4}M_{8}^{-0.81}{\tilde \epsilon}^2 {GM_p\over c^2}.     \label{q68}  
\end{align}

Using eq. (\ref{q67}) we can also calculate a typical GW frequency, $\nu_{GW}={2\over P_{orb}}$, where the factor two 
takes into account the quadrupole character of GW emission. We have
from eq. (\ref{q67})
\begin{align} 
\nu_{GW}={1\over \pi}\sqrt{GM_p\over c_0^3f(e)^3}=3.3\cdot 10^{-5}
\alpha_{r}^{-15/8}\alpha_{s}^{-1/2}{({\mu\over 0.1})}^{-3/8}M_{8}^{0.21}{\tilde \epsilon}^{-3}{f(e)}^{3/2}Hz.   \label{q69}  
\end{align} 

We plot several curves showing the dependencies of $e$ on $a$ 
and $e$ on $\nu_{GW}$ for several values of $\tilde \epsilon$
setting any other parameters in the last expression on r.h.s of eq. (\ref{q68})
to their nominal values  in Fig. \ref{ea}. 
The range of shown semi-major axes
is bounded from below by the condition $r_p > r_{mb}={4GM_p\over c^2}$. Note that the condition $r_c=r_{mb}$ describes correctly the boundary between orbits having periastra and the plunging orbits only when eccentricity is close to unity. We still use it, nevertheless, for any eccentricity, since in any case our oversimplified approach can be used for qualitative purposes only. 

One can see from this Fig. that the formation of substantial eccentricities for values of $a$ comparable with gravitational radius, or even the direct black holes collision are possible when $\tilde \epsilon$ is close enough to unity. But, as we have mentioned above this conclusion has only qualitative character. For
quantitative conclusions a calculation based on full General Relativity is needed, see e.g. \cite{Cutler},
\cite{Flanagan}, \cite{Mino} and references therein. Also,
in all shown cases eccentricities are of the order of
$\sim 0.1-0.3$ in the range of typical GW frequencies available
for LISA $\nu_{GW} > ~ 10^{-5}Hz$, see e.g. \cite{Grish}. We can rewrite eq. (\ref{q63a}) as
\begin{align} 
q_{-2}\approx 4.5 \alpha_{r}^{-3/8}\alpha_s^{1/2}{({0.1\over \mu})}^{3\over 8}M_{8}^{0.24}{\tilde \epsilon}^{-3/8}.   
\label{q70}  
\end{align} 
Equation (\ref{q70}) tells that the condition $\tilde \epsilon 
\approx 1$ correspond to a particular relatively large
mass ratio when all other parameters of the problem are fixed, see also eqns (\ref{q64}) and
(\ref{q65}).
\begin{figure*}
    \includegraphics[width=0.5\linewidth]{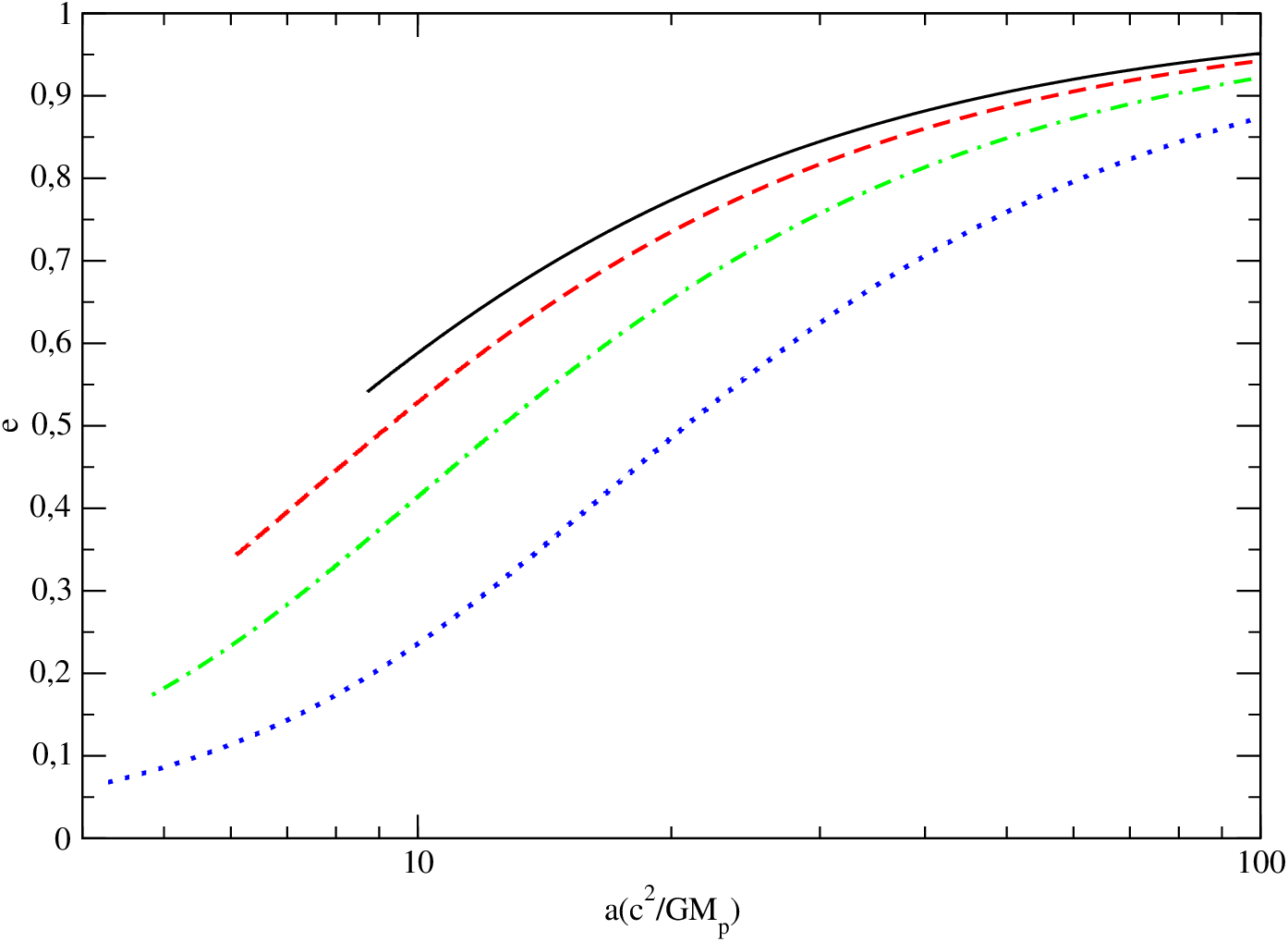}\par 
    \includegraphics[width=0.5\linewidth]{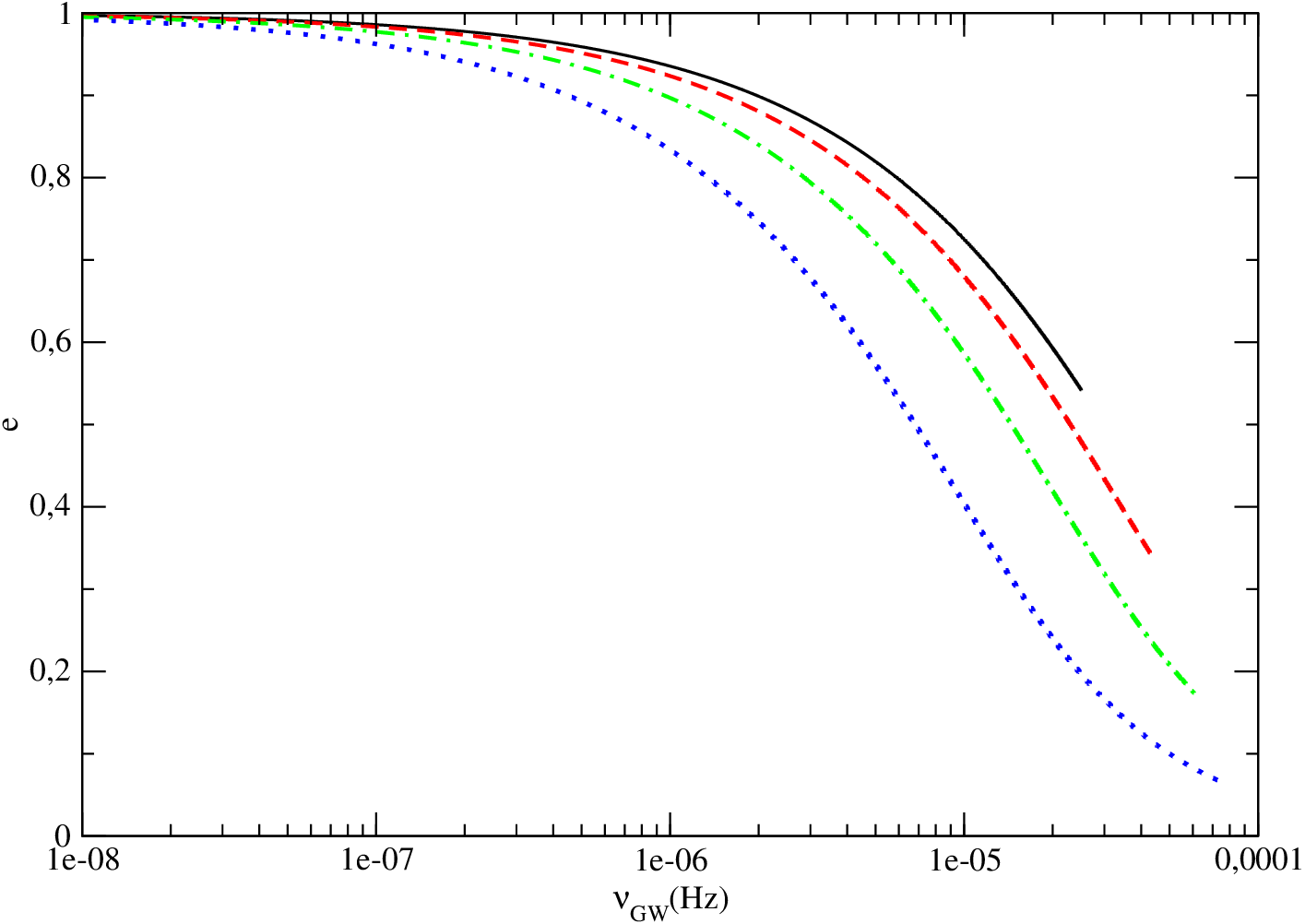}\par 
\caption{Top panel. We show the dependency (\ref{q67}) calculated for $c_0=7.2{\tilde \epsilon}^2{GM_p\over c^2}$ corresponding to the nominal values of all quantities in the
second expression on r.h.s. of eq. (\ref{q68}) on the plane $(({c^2 \over GM_p})a, e)$. The solid, dashed, dot dashed and dotted curves correspond to $\tilde \epsilon=1.1$, $1.2$, $1.4$ and $1.8$, respectively. Bottom panel. The dependency of eccentricity 
on the typical frequency of GW emission calculated according
to eq. (\ref{q69}). Curve of the same type as in left panel correspond to the same value of $\tilde \epsilon$.}
\label{ea}
\end{figure*}}

\bigskip
\section{Conclusions and Discussion}
\label{diss}
\bigskip
In this Paper we consider a secular evolution of a gravitationally bound binary black hole with unequal mass ratio in a center of non-spherical Nuclear Star Cluster (NSC) around the component with a larger mass (primary). The semi-major axis of the binary is supposed to correspond to the radius, where an initial mass of the cluster is equal to the mass of the smaller component (secondary), which is smaller than the radius of influence of the primary. At this radius the evolution of the semi-major axis due to dynamical friction is supposed to be stalled and the stellar mass
in the vicinity of the orbit is supposed to be quickly dispersed by the perturbing action of the secondary. Assuming that these model assumptions are
valid we consider the secular orbital evolution due to tidal forces arising from the non-spherical gravitational field of NSC, which are treated
in the usual quadrupole approximation. We note that by rotation of a coordinate system we can always bring the tidal potential to the form, where only
$m=0$ and $m=2$ terms in its decomposition over spherical harmonics remain. The $m=0$ term is responsible for the usual ZLK secular dynamics, while the secular dynamics due to $m=2$ terms has not, to the best of our knowledge, been explored so far
in the context of SBBH orbital evolution. Note, however, that in the case of stellar mass binaries evolving
in an external field of a star cluster a similar problem was considered by \cite{Petr},
where some numerical calculations and a qualitative analysis were made with the similar conclusions on the possibility of reaching very large values of eccentricity at sufficiently large times. Also, the secular dynamics of stars in a triaxial stellar cluster around a supermassive black hole was discussed in \cite{MV11}, where the possibility of reaching of very large eccentricities was investigated by analytic and numerical means. 

Since in the course of the evolution determined by $m=2$ terms all three components of angular momentum are not conserved, it can potentially lead to the formation of highly eccentric orbits from the orbits with a small or moderate eccentricity, which can, in its turn, promote to an efficient evolution of semi-major axis due to GW emission. Therefore, we consider this evolution in some detail using analytic and numerical means. 

Since the set of equations describing the secular dynamics under the influence of the $m=2$ terms cannot be analytically integrated even when the simplest 
case when only the influence of these terms is taken into account we have to consider some simplifying assumption. The main technical 
assumption made in this work is that there is a moment of time $t=t_0$ in the course of the evolution of our system when eccentricity $e_0$ and inclination to the symmetry plane $i_0$ are small. It then follows that it is possible to construct a singular solution, which is  similar to solutions describing ideal shocks in hydrodynamics in that sense that it also contains discontinuities.  This solution is formally valid in the limit $i_0\rightarrow 0$. It is shown that this solution describes periodic cycles of the evolution. Within every cycle there are two equal time periods corresponding to inclination $i$ being either infinitesimally close to $0$, or to $\pi$. These periods are separated  by two moments of time corresponding to the transitions of inclination from $0$ to $\pi$ and back. During these transitions the eccentricity $e$ formally reaches unity. We find time period of such a solution assuming that $e_0$ is small and show that it is fully characterized by $e_0$ and a value of the nodal angle at the moment $t_0$, $\omega_0$. 

When a small, but finite $i_0$ is considered the transitions of $i$ from $0$ to $\pi$ and back are smeared to a small, but finite periods of time.  We call these periods as {{"sharp jumps" 
}}Also, the eccentricity $e$ reaches a large, but a finite value, which happens at the
moment of time close to the moment of time when $i=\pi/2$. 
We calculate the minimal values of $\epsilon=\sqrt{1-e^2}$, $\epsilon^{\mp}_{min}$, 
where $(-)$ corresponds to the transition of $i$ from $0$ to $\pi$ and $(+)$ corresponds to the opposite transition,  
and show, that they are proportional to trigonometric functions of $\omega_0$ and inversely proportional to $i_0$. Thus, the analytic
theory shows that eccentricity can be made as large as necessary provided that either $i_0$ is chosen to be sufficiently small or $\omega_0$ is equal 
to some special values.

When next order corrections to our asymptotic solution or other factors causing a secular evolution of the orbital elements are considered $i_0$, $e_0$ and
$\omega_0$ are slowly changing from one cycle to another. This also results in a slow evolution of $\epsilon^{\mp}_{min}$, which can, accordingly, reach
other values than the ones corresponding to the initial cycle when a large, in comparison with the period of the cycle time is considered. In this
Paper we consider analytically the most important relativistic factor - the Einstein precession of the apsidal line leaving an analytic treatment of the next order corrections and other factors for a possible future work. It is shown that the role of Einstein precession is determined by the parameter $\beta_{E}$ defined as the ratio of characteristic time of the Einstein precession to that of the secular evolution due to the tidal effects. Based on the law of conservation of Hamiltonian during the secular evolution we argue that a change of $e_0$ on the long timescale is expected to be smaller than  those of $\omega_0$ and $i_0$ and show that the minimal possible value of $\epsilon^{\mp}_{min}$ should be $\beta_{E}/10$. It is then shown that the effect of a 
non zero, but sufficiently small $\beta_{E}$ can be described as a map between $i_0$ and $\omega_0$ corresponding to subsequent cycles. This map has 
the property that either $\epsilon^{-}$ always grows and $\epsilon^{+}$ always decreases or vice versa in the course of the evolution.

The analytic results are supplemented by the results based on numerical solution of the dynamical evolution. First, we show for a purely
Newtonian problem, that when the contribution of $m=0$ term to the full Hamiltonian isn't too large the evolution is qualitatively the same as described above. Next, we consider a large number of cycles taking, for definiteness, approximately $100$ cycles and evaluate numerically the minimal $\epsilon$ over the whole
computational time, $\epsilon_{min}$.  We use $e_{0}$ for all runs and different values of $\omega_0$, $i_{0}$ and $\beta_{E}$. It is shown numerically that
the  averaged over $\omega_{0}$ value of  $\epsilon_{min}$ is close to $\beta_{E}/10$ when $\beta_{E} > 0.05$, for smaller value of $\beta_{E}$ it is
close to constant. This result in full agreement with our theoretical findings, the constant value of 
$\epsilon_{min}$ at small values of $\beta_{E}$ 
is interpreted as being determined by the evolution of $\omega_{0}$ and $i_0$ due to the next order corrections.

We use these results to estimate a possibility of an efficient capture of the secondary having initially an orbit with a small eccentricity $\sim 0.1$
onto an orbit with semi-major axis strongly decreasing due to GW emission in a time corresponding to $100$ cycles of the 
secular evolution. As a criterion of such strong evolution we choose the condition that the semi-major axis should decrease by order of itself per one
orbital period. We use an empirical relation between the size of NSC and the primary mass suggested from observations for early type galaxies, 
the assumption that the radius of influence is the same by order of magnitude as the size and the standard Young profile for the distribution of stellar density inside the sphere of influence. We estimate that for typical masses of the primary and total mass of NSC in the non-spherical component order of
ten per cent the mass ratio should be larger than $0.04$ to fulfill our criterion of the efficient capture. The capture time is estimated to be smaller
than several billion years corresponding to $100$ cycles of the secular evolution for { the assumed typical value of the radius of influence
$r_{infl} \sim 20pc$.  When $r_{infl}$ is ten times smaller, the evolution time is expected to be $\sim 10^{8}yr$, see eq. (\ref{q58}).} Thus, the considered  mechanism provides
an alternative for a solution of the final parsec problem. 

{{ Our estimates of the maximal $\epsilon$, 
and maximal mass ratio $q$ and the primary mass $M_p$ required for the the efficient circularization due to our effect are given by eqns (\ref{q63}), (\ref{q64}) and (\ref{q65}), respectively. It is interesting to note that they contain rather small
powers of the parameters of the problem on r.h.s. This is due to the fact these estimates were obtained by consideration of  relative importance of two competing 
effects. While in order to have a relatively fast secular orbital evolution due to the non-sphericity of NSC larger 
semi-major axes are preferable, in order to have a faster evolution due to GW emission
we need to consider smaller $a$. As a consequence the role of these process is somewhat compensated in the resulting expressions.}}

Assuming that the distribution of the secondary orbits over $i_0$ is uniform a number of systems where our analytical 
theory is directly applicable is quite small, being proportional to $i_0^2$. However, 
our numerical results suggest the values of $\epsilon$ as small as the ones corresponding to small $i_0$ can
be obtained for $i_{0}$ as large as $\pi/2$ when the sufficiently large evolution time is considered, see   Fig. \ref{Fig9}.
A possible explanation of this result is that over a large time there are transient periods of the evolution with 
sufficiently small minimal values of inclination, where the theory described above is qualitatively valid. 
This is illustrated in Fig. \ref{Fig10}. Note that similar results have also been obtained by \cite{Petr},
who considered a long term evolution of a stellar mass binary under the action of a similar, but different perturbing 
potential.

Contrary to the usual scenarios of a SMBBH orbital evolution due to dynamical hardening 
our mechanism leads to a prompt formation of a binary evolving due to GW emission with its orbital periastron comparable
to gravitational radius of the larger component. { For certain parameters of the problem, its eccentricity may remain 
to be appreciable all the way down to the merger, or even the direct black hole collision is possible during the stage of orbital 
evolution caused by GW emission}. Therefore, binaries formed by this mechanism
may be distinguished from the standard ones by a form of gravitational wave signal. In 
this connection it is also interesting to note that the standard model of the activity 
of OJ 287 implies the existence of a binary with similar parameters.

Going back to the Newtonian problem it is interesting to stress the differences of the effect considered above from
the usual ZLK case. One striking difference is that in the case of ZLK the projection of angular momentum onto
the axis perpendicular to the symmetry plane $\propto \cos (i)$ is conserved, and, therefore, the largest values 
of $e$ (the smallest values of $\epsilon$ over an evolutionary cycle correspond to $i\le \sqrt{\sin^{-1} ({2\over 5})}$, 
while in our case the largest values are reached when $i \sim \pi/2$. Another difference is that in the former case the additional integral of motion significantly constraints the smallest values of $\epsilon$, which can be obtained\footnote{ Using eqns (20) of IPS 
(Note a misprint in denominator of the first equation, there should be $e_{1}^2$ instead of $e_1$.), setting there
$\kappa=0$ and assuming, for simplicity, that $\epsilon_1=\sqrt{1-e_1^2}$ and $e_{0}$ are small, we easily obtain
the minimal eccentricity $\epsilon_{1}\approx \sqrt{5\over 8}\cos (i_0)$, where $i_0$ as before 
corresponds to the minimal eccentricity $e_0$. Thus, contrary to our case small values of $\epsilon$ can only 
be obtained for $i_0 \approx \pi/2$. In this approximation the angle $i_1$ corresponding to the smallest $\epsilon$
is approximately equal to $ \sqrt{\sin^{-1} ({2\over 5})(1-e_{0}^2)}$.}.

There are many possible ways of extension of the results reported in this work. One can consider the evolution of the primary
due to dynamical friction down to the radius, where it is assumed to be stalled in stellar cluster with a broken spherical
symmetry and inner cusp, see e.g. \cite{Ses} for an appropriate investigation, but in the case of a spherical initial stellar distribution. Other perturbing factors, such as e.g. the influence of gravitational potential of the stars in the vicinity of the orbit (see e.g \cite{IPS} and \cite{Doug}) 
should be considered. The general 
perturbing potentials having both $m=0$ and $m=2$ terms should be studied as well
as the influence of the neighboring stars. { Also,
the evolution of orbital elements due to GW emission
should definitely be treated in a more accurate way, which can be based on the post-Newtonian expansion of equations of motion (e.g. \cite{Blanch}, \cite{Gopak} and references
therein) or using a relativistic approximation scheme based on a small mass ratio of the black holes (e.g. \cite{Cutler},
\cite{Flanagan}, \cite{Mino} and references therein).}
It is clear that the general consideration 
can only be done by numerical means. This is left for a future work.

Finally, we would like to note that our results can be potentially applied in other astrophysical settings. For example, 
one can consider a binary star or an exoplanetary system inside a massive non-spherical protostellar cloud. Tidal forces
acting from the cloud may influence the secular dynamics of a considered system in a way similar to what is discussed above.
 
\acknowledgements 

We are grateful to A. I.Neishtadt, E. V. Polyachenko, S. V. Pilipenko, V. V. Sidorenko and E. A. Vasiliev
for important comments and discussions. 

\begin{appendix}

\section{The evolution of $\epsilon$ in case when $\phi \ne \pi/2$}
\label{phi}

\begin{figure*}
    \includegraphics[width=0.5\linewidth]{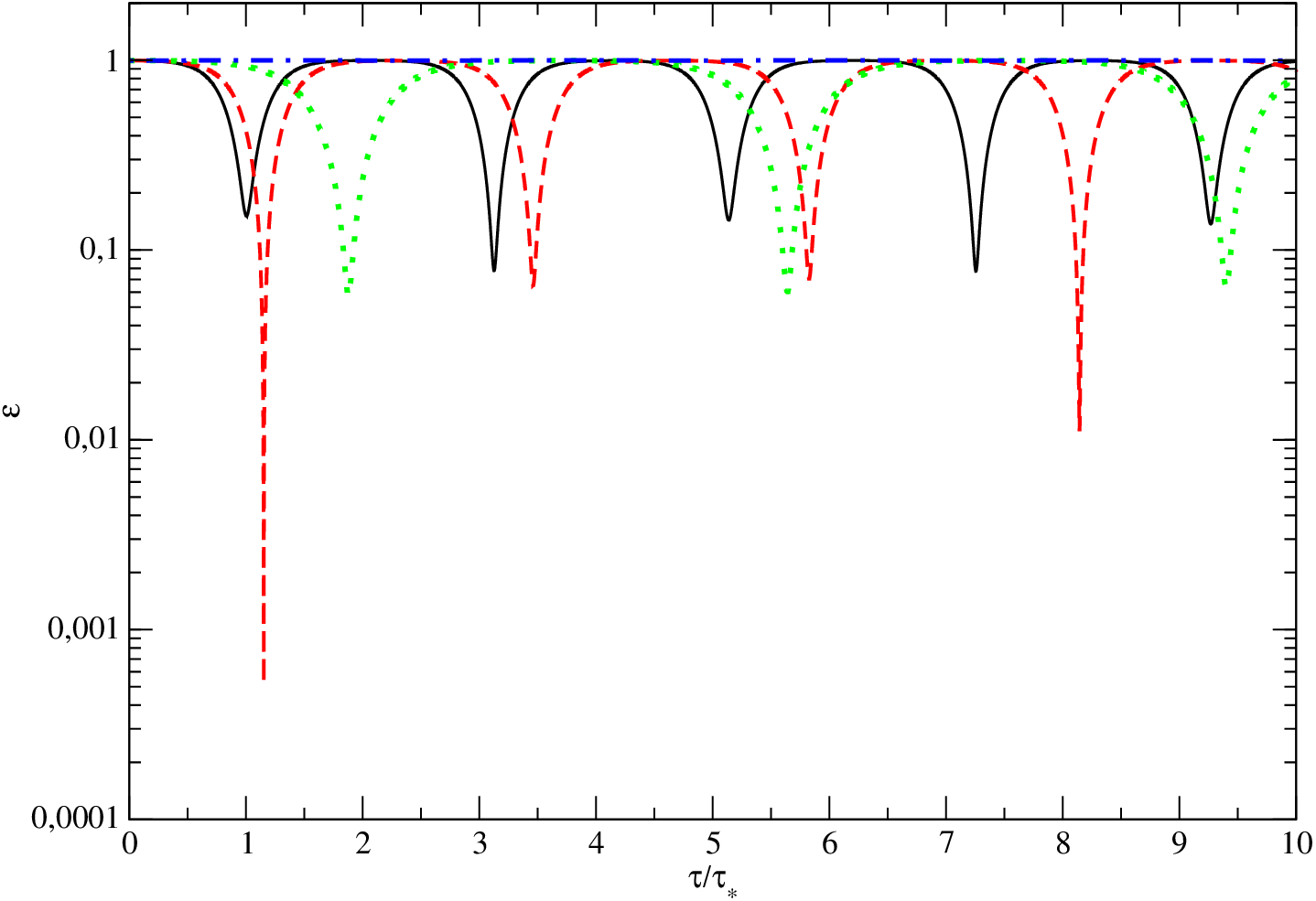}\par 
\caption{The numerically obtained dependencies of $\epsilon$ on time for different values of $\phi$.
The initial values of  $i$, $e$ and $\omega$ are the same as the ones used in Figs. \ref{Fig2} and \ref{Fig3}.
The solid, dashed, dotted and dot dashed curves correspond to $\phi=\pi/2$, $3\pi/8$, $\pi/4$ and $\pi/8$,
respectively.}
\label{Fig5}
\end{figure*}

For simplicity, we do not provide the same asymptotic analytic treatment for the general case, when
$\phi \ne \pi/2$ and both sets of dynamical equations (\ref{q8}) and (\ref{q9a}-\ref{q9b}) contribute
to the evolution of our system according to the rule (\ref{q4}).
 However, we have made
a set of additional numerical calculations choosing the same 'standard' initial values of the dynamical variables as in Figs \ref{Fig2} and \ref{Fig3}, but varying $\phi$ from $0$ to $\pi/2$. The result of calculations of $\epsilon$ for these cases is shown in
Fig. \ref{Fig5}.
One can see from this Figure that the dependency of the values of minimal $\epsilon$ on $\phi$ is
non-monotonic. Note a very small value of the minimal $\epsilon$ in case when $\phi=3\pi/8$. Even in
the case when $\phi=\pi/4$ and, accordingly, the terms with $m=0$ and $m=2$ give the same contribution
to the perturbing potential the values of minimal  $\epsilon$ are close to, but smaller than in the
case $\phi=\pi/2$, where only the $m=2$ term is taken into account. Only in the case when $\phi=\pi/8$ 
variations of $\epsilon$ are insignificant. Also note that the period of time passed between successive
minima of $\epsilon$ is comparable to the case $\phi=\pi/2$. We conclude that our results discussed above
can be extended to the general case for the purpose of making of qualitative estimate unless the contribution of the $m=2$ term isn't small. 

\section{The iterative map describing the influence of Einstein precession}
\label{map}

In order to find the explicit form of the rule of change of the 'initial value' of nodal angle and inclination
we should relate the constant $C$ to the expressions (\ref{q18b}- \ref{q23}). Let us first consider the first transition
when $\kappa=-1$. In this case it follows from eq. (\ref{q18b}), where we set $e=1$ and $\epsilon=0$ in 
the denominator on r.h.s.,
that
\begin{align} 
\Delta_{1}=2\epsilon^{-}_{min}e_{0}^{2/5}\cot (\psi_i)T,
\label{q32}  
\end{align}  
where we use the fact that $\epsilon \approx \epsilon^{-}_{min}|T|$ in the limit $|T|\rightarrow \infty$,
and $i=0$ when $T$ is negative and $i=1$ otherwise. We also remind that $\psi_{i}=\omega_{i}-{\pi\over 4}$. 
The expression (\ref{q31}) provides the link between
the values of the constant in front of $T$ corresponding to the negative and positive values of this variable.
We obtain from eqns (\ref{q31}), (\ref{q32}) and the definition of $\beta_{*}={\beta_E\over 10\epsilon^{-}_{min}}$ that
\begin{align} 
2\epsilon^{-}_{min}e_{0}^{2/5}\cot (\psi_0)+{\beta_{E}\over 5\epsilon^{-}_{min}} = 
2\epsilon^{-}_{min}e_{0}^{2/5}\cot (\psi_1).
\label{q33}  
\end{align} 
The additional requirement is that a value of $\epsilon^{-}_{min}$ should not depend on whether we calculate
it using $(\omega_{0}, i_{0})$ or $(\omega_{1}, i_{1})$. Dividing eq. (\ref{q33}) by $2\epsilon^{-}_{min}e_{0}^{2/5}$ 
and using eq. (\ref{q25a}) with $\kappa=-1$ we obtain
\begin{align} 
\cot (\psi_0)+{\beta_{E}\over 10i_{0}^{2}\sin^{2}(\psi_0)} = 
\cot (\psi_1),
\label{q34}  
\end{align} 
and the additional requirement leads to
\begin{align} 
i_{0}|\sin(\psi_{0})|=i_{1}|\sin(\psi_{1})|.
\label{q35}  
\end{align} 
It is obvious that when $\beta_{E}=0$ $\psi_{1}=\psi_{0}$ and $i_{1}=i_{0}$. The expressions (\ref{q34}) and
(\ref{q35}) define the change of the constants $\omega_{0}$ and $i_{0}$ under the influence of Einstein 
apsidal precession when the inclination angle changes from $\sim 0$ to $\sim \pi$.

In order to consider the second transition of $i$ from $\sim \pi$ to $\sim 0$ we proceed in a similar
way. We use eq. (\ref{q18c}) in the limit $e\rightarrow 1$ and $\epsilon \rightarrow 0$ and substitute 
there the expression for $C_{\omega}^{+}$ to obtain
\begin{align} 
\Delta_{1}=-2\epsilon^{+}_{min}e_{0}^{2/5}\tan (\psi_i)T,
\label{q36}  
\end{align}  
which is different from eq. (\ref{q32}) in two ways. Namely, we have tangent of $\psi_i$ instead of
cotangent in eq. (\ref{q32}) and there is a minus sign in the front of expression. It then follows from eq. (\ref{q25a}) 
with $\kappa=+1$ and eq. (\ref{q31}) 
that the transition from $(\omega_{1}, i_{1})$ or $(\omega_{2}, i_{2})$ should have a form similar to (\ref{q34}) 
and (\ref{q35}), but with $\tan (\psi_i)$ instead of $\cot (\psi_i)$, $\cos (\psi_i)$ instead of $\sin (\psi_i)$ and
the sign minus in the front of $\beta_{E}$. We have
\begin{align} 
\tan (\psi_1)-{\beta_{E}\over 10i_{1}^{2}\cos^{2}(\psi_1)} = 
\tan (\psi_2),
\label{q37}  
\end{align} 
and the additional requirement leads to
\begin{align} 
i_{1}|\cos (\psi_{1})|=i_{2}|\cos (\psi_{2})|.
\label{q38}  
\end{align} 

The expressions (\ref{q34}), (\ref{q35}), (\ref{q37}) and (\ref{q38}) provide an analytic map between initial values $(i_0, \omega_{0})$ and the values of the same quantities in the end of the cycle,  $(i_2, \omega_{2})$, which is non-trivial when
the Einstein precession takes place. Considering many iterations of this map we can see how the Einstein precession 
changes values of $\epsilon_{min}^{\kappa}$ and, accordingly, the probability of the formation of a close binary due to GW emission,
on timescales $\gg 4t_{*}$. Note that it follows from eq. (\ref{q34}) that the map is singular when $\omega_0={\pi/4}$ and,
accordingly, $\psi_0=0$, see eq. (\ref{q18a}). The map also cannot accurately describe the considered process when  
$\omega_0=-{\pi/4}$ corresponding to $\psi_0=-{\pi/2}$. In the latter case eq. (\ref{q32}) tells that $\Delta_1$ is formally
equal to zero when $T\rightarrow -\infty$. In this situation one should consider next order terms in the expansion in powers
of $\epsilon$ in eq. (\ref{q18b}). We do not treat these special cases in this Paper assuming that $\omega_{0}$ is sufficiently 
different from $\pm {\pi/4}$.  

\begin{figure*}
    \includegraphics[width=0.5\linewidth]{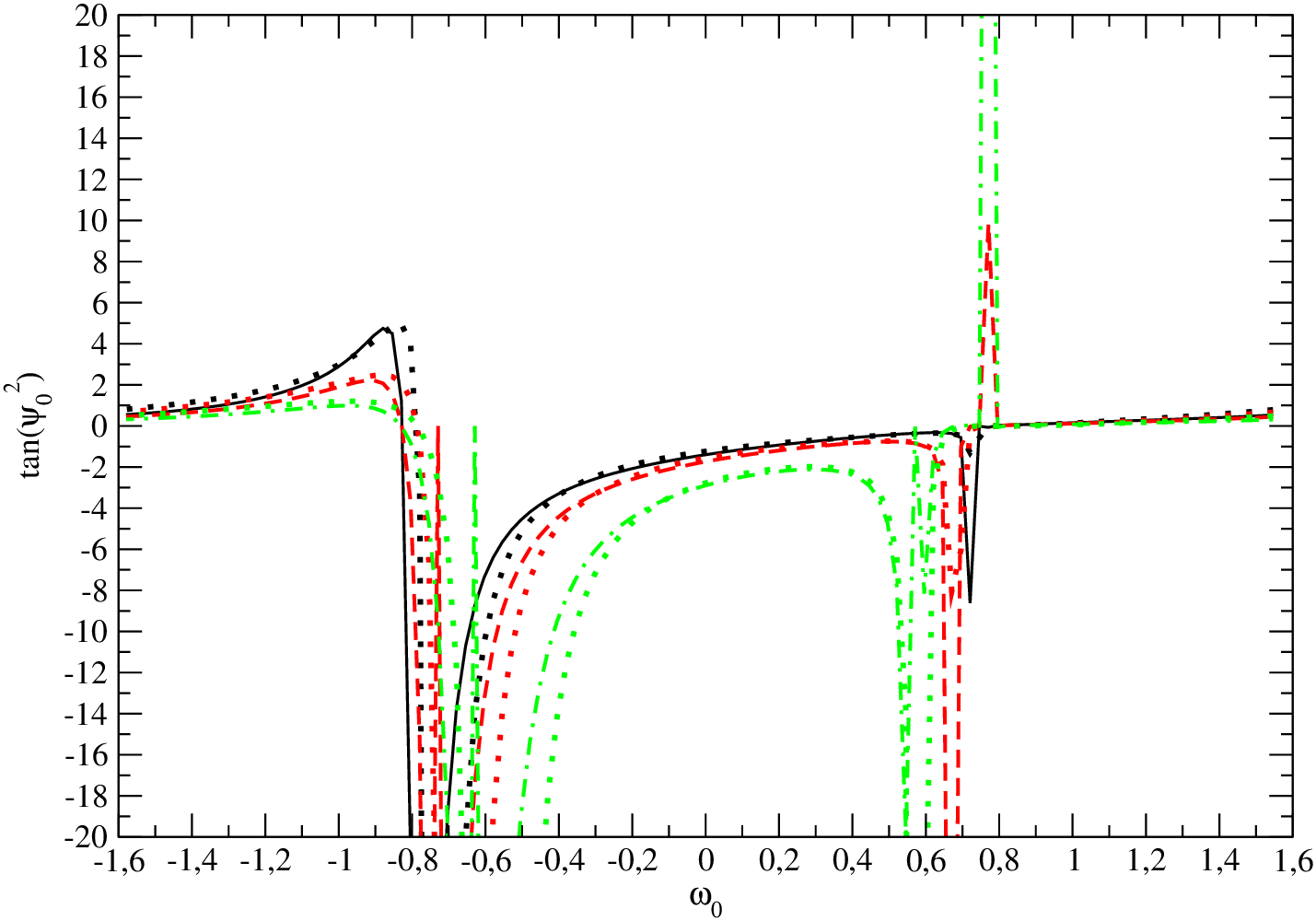}\par 
\caption{We show $\tan(\psi_2)$ numerically obtained using  eqns (\ref{q34}), (\ref{q35}), (\ref{q37}) and (\ref{q38})
after one iteration as a function of $\omega_0$ together with the corresponding results obtained by numerical solution
of our equations of motion for the time period $\tau=4\tau_{*}$. $e_0=0.1$, $i_0=0.1$ 
for all shown cases. Solid, dashed and dot-dashed lines represent the results of the numerical solution for $\beta_E=5\cdot 10^{-3}$, $10^{-2}$ and $2\cdot 10^{-2}$, respectively, while the dotted lines are obtained using the map.}
\label{Fig6}
\end{figure*}

The results of comparison of our map with numerical solution of our equation of motion in the time interval $\tau < 4\tau_{*}$ is
shown in Fig. \ref{Fig6}, for different values of $\omega_{0}$ in the range $-{\pi\over 2} < \omega_{0} < {\pi\over 2}$ and 
fixed values of other parameters of the problem. One can see that the map gives quite good agreement with the fully numerical
results apart from narrow regions close to $|\omega_{0}|={\pi\over 4}$, where both approaches strongly deviate.

\begin{figure*}
    \includegraphics[width=0.5\linewidth]{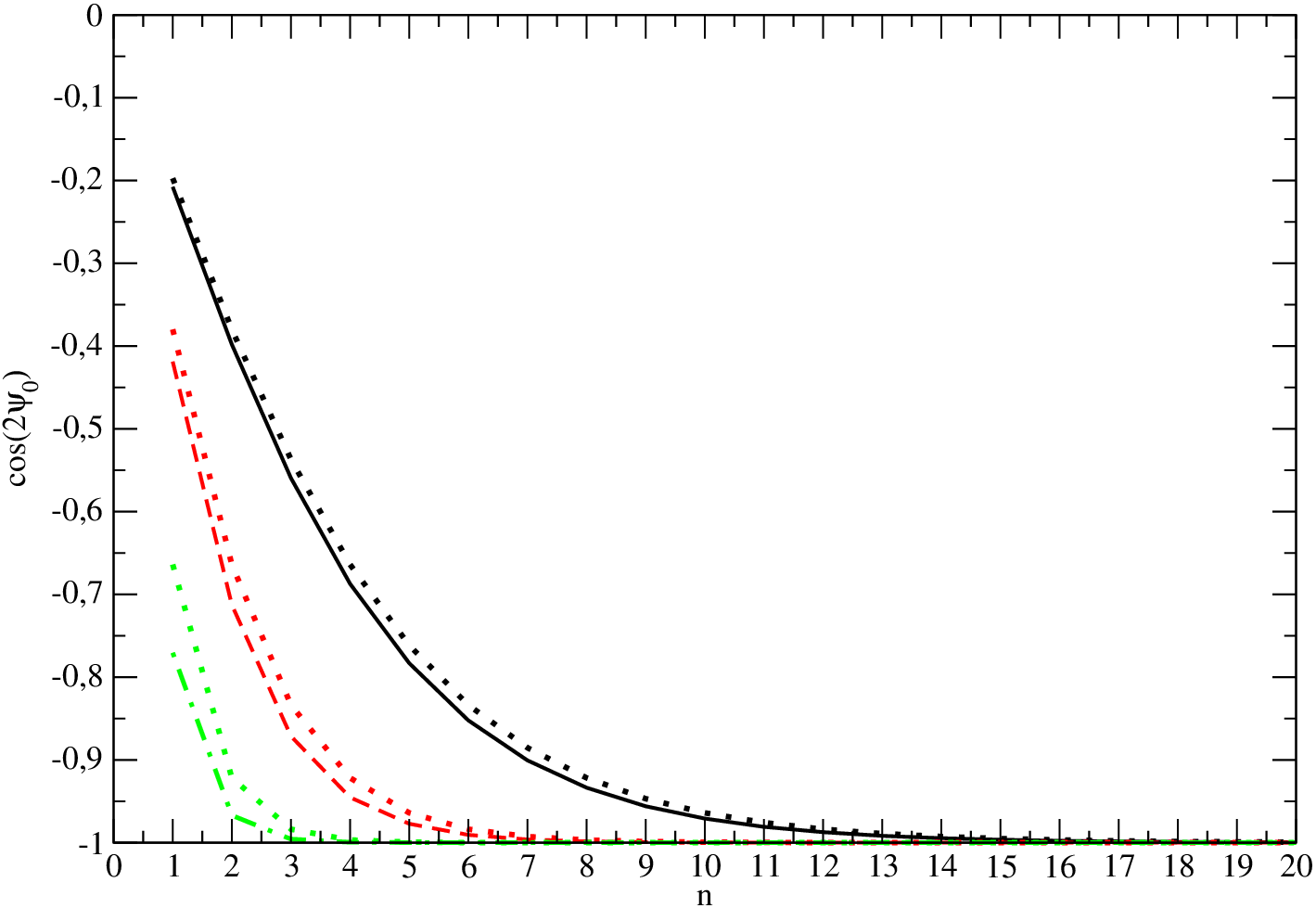}\par 
    \includegraphics[width=0.5\linewidth]{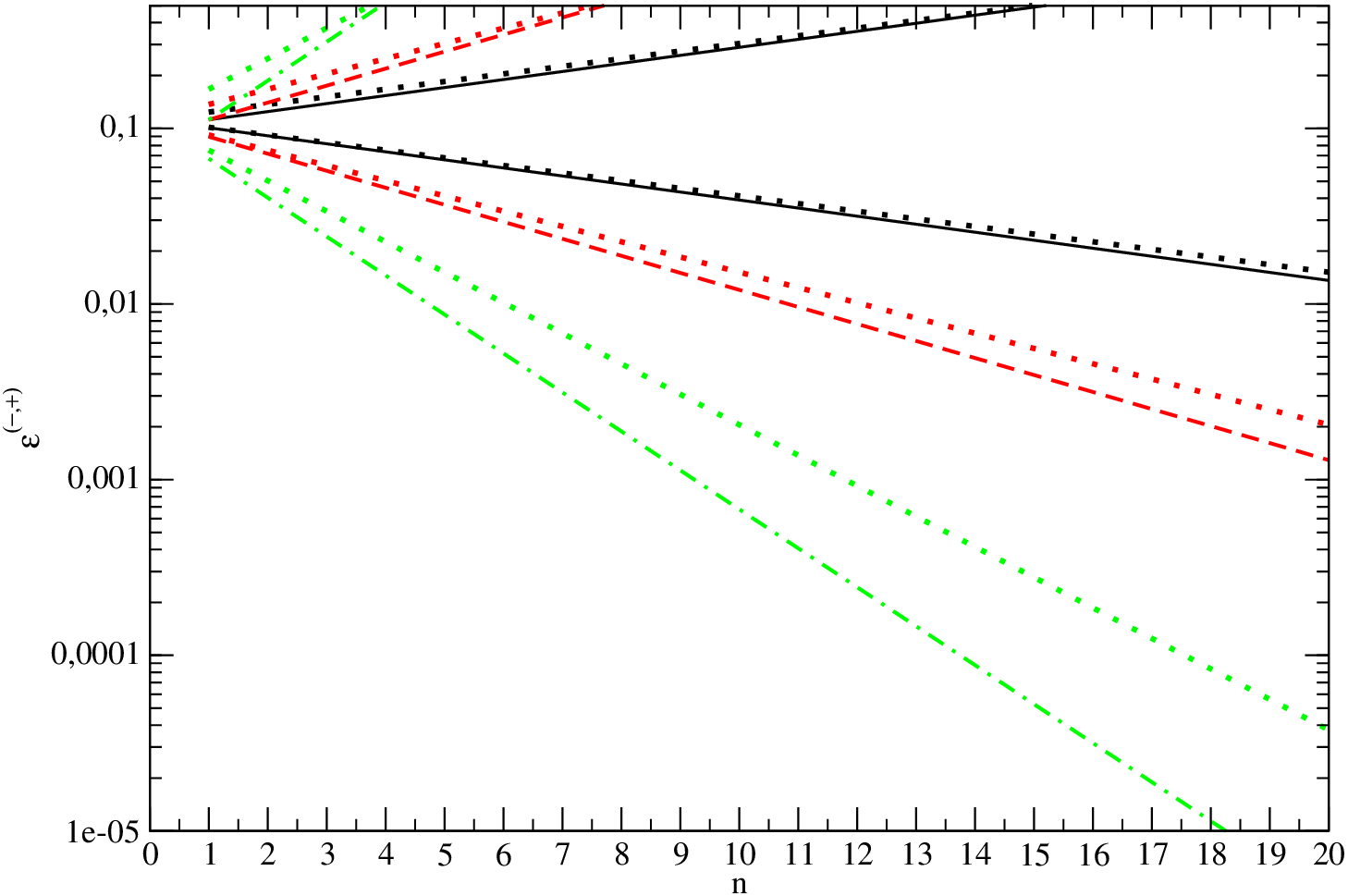}\par 
\caption{Top panel. 
The result of comparison of the map defined by eqns
(\ref{q34}), (\ref{q35}), (\ref{q37})
and (\ref{q38}) with our
analytic expressions. We show $\cos (2\psi_{0})$ numerically calculated with the help of the map as a function $n$,
for $e_0=0.1$, $\psi_{0}(0)={3\pi\over 4}$, $i_{0}(0)=0.1$ and $\beta_E=5\cdot 10^{-3}$, $10^{-2}$ and $2\cdot 10^{-2}$ as
solid, dashed and dot-dashed lines, respectively. Since $\sin (2\psi_{0}(0)) < 0$ for this case the appropriate analytic
expressions are given by eq. (\ref{q44}). These are shown by dotted lines. Bottom panel. The evolution of $\epsilon^{\mp}$ 
for this case. The styles of the lines are for the same parameters as in top panel. The  dotted curves are calculated according
to eq. (\ref{q46}).}
\label{Fig7}
\end{figure*} 

It is instructive to simplify the expressions (\ref{q34}), (\ref{q35}), (\ref{q37}) and (\ref{q38}) assuming that changes of 
the angles $\psi$ and $i$ during one cycle, $\Delta \psi=\psi_2-\psi_1$ and $\Delta i=i_2-i_0$ are small in comparison with
the initial values. It is then straightforward to obtain
\begin{align} 
\Delta \psi=-{\beta_{E}\over 5i_{0}^2}, \quad  \Delta i= -i_{0}\cot (\psi_0){\Delta \Psi \over 2}.
\label{q39}  
\end{align}  
From the conditions $|\Delta \psi | <1 $ and $|\Delta i|/i_0 < 1$ we obviously have
\begin{align} 
\beta_{E} < \min (1, |\tan (\psi_0)|)5i_{0}^2.
\label{q40}  
\end{align} 
The condition (\ref{q40}) tells again that our approach is invalid when $\omega_{0} \approx \pi/4$, and, accordingly, 
$\psi_0 \approx 0$.

Assuming that this condition holds we can extend the map (\ref{q39}) to many cycles by reducing it to ordinary differential
equations according to the rule $\Delta \psi \rightarrow {d\psi_0\over dn}$ and $\Delta i \rightarrow {di_{0}\over dn}$, where
$n$ is a number of cycles \footnote{Remembering that the time period of one cycle is $4\tau_{*}$, it possible to use the time
as an independent variable instead of $n$ with the help of substitution $n={\tau \over 4\tau_{*}}$.}. Proceeding in this way 
we see that the second relation in eq. (\ref{q39}) can be immediately integrated and the result can be substituted in the first
relation to give
\begin{align} 
{d\psi_0\over dn}=-{\beta_E\over {i_{0}(0)}^{2}|\sin (2\psi_{0}(0))|}\sin (2\psi_{0}), \quad 
i_{0}=i_{0}(0)\sqrt{{|\sin (2\psi_{0}(0))| \over |\sin (2\psi_{0})|}},
\label{q41}  
\end{align}  
where $i_{0}(0)$ and $\psi_{0}(0)$ are values of our variables in the beginning of the evolution corresponding to
$n=0$. The first equation can be easily integrated with the result
\begin{align} 
\cos 2\psi_0={\cot^2(\psi_0(0))\exp ({4\over 5}\phi_E)-1\over \cot^2(\psi_0(0))\exp ({4\over 5}\phi_E)+1} \quad
\phi_{E}= {\beta_E n\over {i_0(0)}^2\sin (2\psi_{0}(0))}.
\label{q42}  
\end{align} 
In the asymptotic limit of large  $n$  we obtain from eq. (\ref{q42}) and the second expression in eq. (\ref{q41})
\begin{align} 
\sin \psi_0 \approx |\tan (\psi_0(0))|\exp (-{2\over 5}\phi_{E}), \quad \cos \psi_{0}\approx 1, \quad i_0\approx
i_0(0)|\cos (\psi_0(0))|\exp ({\phi_E\over 5})
\label{q43}  
\end{align} 
when $\sin (2\psi_0(0)) > 0$ and
\begin{align} 
\cos \psi_0 \approx |\cot (\psi_0(0))|\exp (-{2\over 5}|\phi_{E}|), \quad \sin \psi_{0}\approx 1, \quad
i_0\approx i_0(0)|\sin  (\psi_0(0))|\exp ({|\phi_E|\over 5})
\label{q44}  
\end{align} 
in the opposite case.
Finally, we substitute eqns (\ref{q43}) and (\ref{q44}) in eq. (\ref{q25a}) to obtain
\begin{align} 
\epsilon^{-}={i_{0}(0)|\sin \psi_{0}(0)|\over e_{0}^{1/5}}\exp (-{\phi_E\over 5}), \quad 
\epsilon^{+}={i_{0}(0)|\cos \psi_{0}(0)|\over e_{0}^{1/5}}\exp ({\phi_E\over 5}) 
\label{q45}  
\end{align} 
when $\sin (2\psi_0(0)) > 0$ and
\begin{align} 
\epsilon^{-}={i_{0}(0)|\sin \psi_{0}(0)|\over e_{0}^{1/5}}\exp ({|\phi_E|\over 5}), \quad 
\epsilon^{+}={i_{0}(0)|\cos \psi_{0}(0)|\over e_{0}^{1/5}}\exp (-{|\phi_E|\over 5}) 
\label{q46}  
\end{align} 
in the opposite case. 

We compare our analytic results with the corresponding results based on the use of the map in Fig. \ref{Fig7}. 
One can see a very good agreement between both approaches in the case when $\beta_{E}$ is sufficiently small $\sim < 0.01$. 
For larger value of $\beta_{E}=2\cdot 10^{-2}$ the results are still in agreement, but it gets worse as expected.

\end{appendix}

\def\apj{Astrophys.~J}
\def\apjl{Astrophys.~J.,~Lett}
\def\apjs{Astrophys.~J.,~Supplement}
\def\an{Astron.~Nachr}
\def\aap{Astron.~Astrophys}
\def\mnras{Mon.~Not.~R.~Astron.~Soc}
\def\pasp{Publ.~Astron.~Soc.~Pac}
\def\aaps{Astron.~and Astrophys.,~Suppl.~Ser}
\def\apss{Astrophys.~Space.~Sci}
\def\ibvs{Inf.~Bull.~Variable~Stars}
\def\japa{J.~Astrophys.~Astron}
\def\na{New~Astron}
\def\aspproc{Proc.~ASP~conf.~ser.}
\def\aspcs{ASP~Conf.~Ser}
\def\aj{Astron.~J}
\def\actaa{Acta Astron}
\def\araa{Ann.~Rev.~Astron.~Astrophys}
\def\caosp{Contrib.~Astron.~Obs.~Skalnat{\'e}~Pleso}
\def\pasj{Publ.~Astron.~Soc.~Jpn}
\def\memsai{Mem.~Soc.~Astron.~Ital}
\def\astl{Astron.~Letters}
\def\aipproc{Proc.~AIP~conf.~ser.}
\def\physrep{Physics Reports}
\def\sovast{Soviet~Ast.}
\def\planss{Planet.~Space~Sci.} 

\bibliographystyle{apsrev4-2}
\bibliography{hami.bib}

\end{document}